\documentclass[12pt,preprint]{aastex}

\begin{document}

\title{Far-Infrared Galaxies in the Far-Ultraviolet\footnote{Based on
observations with the NASA/ESA Hubble Space Telescope obtained at the
Space Telescope Science Institute, which is operated by the Association
of Universities for Research in Astronomy, Incorporated, under
NASA contract NAS5-26555.}}

\author{Jeffrey D. Goldader}
\affil{Department of Physics and Astronomy, University of Pennsylvania, 
Philadelphia, PA 19104}
\email{jdgoldad@dept.physics.upenn.edu}

\author{Gerhardt Meurer, Timothy M. Heckman, Mark Seibert}
\affil{The Johns Hopkins University, Baltimore, MD 21218}

\author{D. B. Sanders}
\affil{Institute for Astronomy, University of Hawaii, 2680 Woodlawn Drive,
Honolulu, HI 96822}

\author{Daniela Calzetti}
\affil{Space Telescope Science Institute, Baltimore, MD 21218}

\author{Charles C. Steidel}
\affil{Palomar Observatory, California Institute of Technology, Pasadena, 
CA 91125}

\begin{abstract}
In an effort to 
better understand the UV properties of ultraluminous infrared
galaxies (ULIGs), and compare them 
to the rest-frame UV properties of high redshift sub-mm and 
Lyman-break galaxies, we have 
obtained far- and near-UV imaging observations ($\lambda_{eff}=1457\mbox{\AA}$,
$\lambda_{eff}=2364\mbox{\AA}$, respectively) of two luminous infrared galaxies
(LIGs--VV 114 and IC 883) and five ULIGs 
(IRAS 08572+3915, Mrk 273, IRAS 15250+3609, 
Arp 220, and IRAS 19254--7245) using the Hubble Space Telescope.
All the galaxies were detected in both channels.
UV light, both diffuse and from star clusters, can be traced to within
the inner kpc of the dominant near-IR nuclei.  However, in general,
the brightest UV sources are clearly displaced from the 
$I$-band and near-IR peaks by at least hundreds of pc. Further,
only 0.07\%-7.3\% of the total near-UV 
light is projected within the inner 500 pc radius, even though this
is the same region wherein most of the bolometric energy is generated.
All nuclei are highly obscured by dust. 
Even after correction for
dust reddening, the global UV emission fails to account for the total 
bolometric luminosities of these systems by factors of 3--75.
The discrepancy is much worse if only the central regions, where
the bolometric luminosities are generated, are included.
In two cases (VV 114 and IRAS 08572+3915), the merging companion galaxies
are more prominent in the UV than the more IR-luminous member.
While all our galaxies show possible signatures of AGN activity,
only IRAS 19254--7245
yields even a possible detection of an AGN in our UV images.
Simple calculations show that all but one of our galaxies would be expected 
to drop below 
the detection thresholds of, e.g., the Hubble Deep Fields at redshifts
between 1.5 and 3, and we find that $\sim$2 of our 5 ULIGs would be
selected as Extremely Red Objects in this redshift range.
A typical ULIG in our sample would be too faint to be detected at 
high-redshift in the deepest current optical or sub-mm deep surveys.
Only VV 114 has UV luminosity and color similar to Lyman-break galaxies
at $z\sim 3$;
the other galaxies would be too faint and/or red to be selected by 
current surveys.  The low UV brightnesses of our ULIGs mean that they would
not appear as optically-bright (or bright ERO) sub-mm galaxy 
counterparts, though they might be similar to the fainter sub-mm 
galaxy counterparts.
\end{abstract}


\section{Introduction}

A significant challenge in astronomy today is understanding the
relationships between very distant galaxies and nearby galaxies.  Recent
observations have identified many luminous, high-redshift systems, yet
they are faint and very difficult to study in detail.  Are there good
local counterparts for these galaxies?  If so, we may try and 
use the local counterparts to better understand their distant relatives.

New observations, carried out primarily with the Sub-millimeter Common 
User Bolometer Array (SCUBA: Holland et al. 1999) 
on the James Clerk Maxwell Telescope, 
have revealed the presence of a significant population of far-IR (FIR) 
luminous galaxies, most likely at redshifts $z \ga 2$ (``sub-mm galaxies,"
e.g., Smail et al.  1998, Barger et al. 1999a).
These galaxies are apparently responsible for much of the FIR and 
sub-mm extragalactic background radiation, whose energy density is
comparable to that of the integrated optical light of faint
galaxies seen in, e.g., the Hubble Deep Fields (HDFs) 
(e.g., Hauser et al. 1998; Barger, Cowie \& Sanders 1999).
The sub-mm galaxies detected to date are much more luminous than almost
any galaxy in the local universe.
It is possible
that a significant fraction of the star formation in the early universe
occurred in FIR-luminous galaxies, of which the current sample of sub-mm
galaxies are the most luminous.  At present, it is not clear if the
sub-mm galaxies are at all related to galaxies at similar redshifts
discovered by the Lyman break technique, even though the brightest, reddest
Lyman break galaxies (LBGs) may have comparable bolometric luminosities,
after reddening is taken into account \citep{meu99,ade00}.  Most (but not
all) sub-mm galaxies are too faint and/or red to be found with LBG
selection criteria.

For the sub-mm galaxies with known redshifts, the bluest sub-mm observations
(at, e.g., 450 \micron) still fall on the Rayleigh-Jeans side of the
rest-frame FIR peak wavelengths.  Nonetheless,
the limited information on these sub-mm galaxies shows that they appear 
to have overall 
rest-frame FIR/radio spectral 
energy distributions quite similar to those of the ultraluminous 
infrared galaxies discovered in the local universe ($z\la 0.1$).
This has led to increased interest in the nearby infrared galaxies
as possible counterparts.

Ultraluminous infrared galaxies 
\footnote{See \citet{san96} for a review.  $L_{IR}=L(8-1000\micron)$,
computed using the prescription in that paper.}  (ULIGs, with
$L_{IR}>10^{12}L_{\sun}$) were found to be a significant class of
objects upon analysis of the Infrared Astronomical Satellite (IRAS)
all-sky survey \citep{soi87, san88}.  The bolometric luminosities of
ULIGs, the most powerful infrared-luminous galaxies, are similar to
those of optically-selected quasars.  We now understand ULIGs to be a
stage in the merger of two gas-rich spiral galaxies.  During the merger,
the molecular gas is driven inward towards the nuclei, where the gas
serves to fuel a tremendous burst of star formation and, perhaps, AGN
activity as well.  Dust absorbs and re-radiates the light from young
stars and/or the AGN.  Complicating our understanding of ULIGs is the
fact that most of the energetic activity takes place in the inner few
hundred pc, where small angular sizes and almost entirely opaque ISM
prevent us from directly witnessing events occurring in the nuclei
themselves.

The primary goal of our study is to explore the possible connection 
between sub-mm galaxies and ULIGs.
There is substantial circumstantial evidence linking ULIGs and sub-mm 
galaxies, but a direct connection has not been established.  In part, this is
because of observational considerations. The FIR/radio spectral energy
distributions (SEDs) of 
the best-studied
sub-mm galaxies (ERO J164502+4626.4: Dey et al. 1999;
SMM J02399--0136: Ivison et al. 1998; SMM J14011+0252: Ivison et al. 2000;
and SMM J02399--0134: Barger et al. 1999a)
are sampled redward of the rest-frame FIR peak, which should occur
at 60--100 \micron\ if their dust temperatures are similar to those of
the ULIGs.  Optical and near-IR observations sample blueward of
$\sim$0.6 \micron\ in the rest-frame for $z\sim 2.5$.
As a result, we know little or 
nothing of the properties of sub-mm
galaxies from the red optical through $\sim$200 \micron\ in the rest-frame.
And unfortunately, local ULIGs are poorly studied shortward of 
$\sim$4000 \AA.  What morphological comparisons we can make using,
e.g., optical images, are quite limited.
As a result, the strongest links between ULIGs and sub-mm galaxies
are made largely on the basis of three facts.  First, ULIGs and 
sub-mm galaxies
are the most 
highly luminous systems in the far-IR in the local universe and at $z
\ga 1$ respectively, although most sub-mm galaxies detected so far are inferred
to
be still a
factor of few more luminous than ULIGS such as Arp 220 (Barger, Cowie
\&\ Richards 2000). 
Second, ULIGs and sub-mm galaxies also have similar far-IR/sub-mm/radio SEDs.  
Finally, large amounts of molecular gas (e.g., CO) have been detected 
in both classes of objects.  

Local counterparts, even when not exact matches, can offer
important insights on high-{\em z} systems.  For example,
starbursts observed with IUE (e.g., Meurer et al. 1999
and references therein) have SEDs and UV spectral properties like LBGs, 
though the local galaxies are smaller and less luminous; this has been used
to infer that the basic physics (e.g., winds, reddening) of 
these objects are similar.  We hope that observations of ULIGs in
the rest-frame UV, telling us about their morphologies, luminosities,
and colors, might both allow a test of their fitness as counterparts of
sub-mm galaxies, and also allow us to better interpret the rest-frame UV
observations of sub-mm galaxies.

A secondary goal is to test whether ULIGs follow what is known as
the ``IRX-$\beta$" correlation
between the redness of the UV continuum (parameterized by the UV spectral 
slope $\beta$ between $\sim1600\mbox{\AA}$ and $2200\mbox{\AA}$, where 
$f_\lambda \propto \lambda^{\beta}$) and the 
IR-excess (the IRX, defined as the ratio of 
the FIR/UV fluxes\footnote{In this paper, we compute the FIR fluxes according
to the prescription of Helou et al. 1988, including the dust 
temperature dependent
color
correction factor defined in the appendix of that paper.  We note that
Meurer et al. use a constant color correction factor reasonable for
the typical dust temperatures of their galaxies.  The UV flux
is computed as $F_{\lambda}=\lambda \times f_{\lambda}$ in 
the FUV channel, where 
$f_\lambda$ is the observed flux density in erg/cm$^2$/s/$\mbox{\AA}$.}), 
that has been established 
for a sample of local starbursts observed in the 
UV \citep{meu95,meu97,meu99}.  
The correlation is well fit by a foreground dust
shell model, wherein dust surrounding the UV-bright region both
reddens and absorbs the UV light, reprocessing it to FIR wavelengths.
The correlation appears to hold for $L\la 10^{11.5} L_{\odot}$, but
has not been tested at higher luminosities.
If ULIGs also follow this correlation, then we would have more
confidence that rest-frame UV observations of even the most dusty
high-$z$ star-forming galaxies can be used to estimate their total
bolometric luminosities (dominated by the difficult to access
far-infrared) ensuring that these galaxies are included in the
census of star formation in the early universe.
We know that $\beta$ and the IRX correlate with total luminosity
for the Meurer et al. sample \citep{hec98}.
If these trends hold at higher $L$ and higher $z$, then
sub-mm galaxies may be too red to be selected as LBGs, and may perhaps
better be associated with Extremely Red Objects (e.g.\ Ivison et
al.\ 2000, ApJ, 542, 27).

In this paper, we present new rest-frame UV observations of 
seven galaxies with $L_{IR}>10^{11.5}$.
Three ULIGs (VII Zw 31, IRAS 12112+0305, and IRAS 22491--1808) 
were detected in the UV 
using the Faint Object Camera on the pre-COSTAR Hubble Space Telescope (HST)
by Trentham, Kormendy, \& Sanders (1999, hereafter TKS).  Though their 
photometric precision was limited by the spherical aberration and 
high backgrounds, the detections 
were good evidence that our program was in fact feasible.
More recently, ground-based images in the $U^{\prime}$ band 
($\lambda=3410\mbox{\AA}$, $\Delta\lambda=320\mbox{\AA}$) were obtained for
many ULIGs (including three of our galaxies, IRAS 08572+3915, Mrk 273, and
IRAS 15250+3609, plus two galaxies, IRAS 12112+0305 and IRAS 22491--1808, 
from TKS) 
by \citet{sur00a}.

This paper concentrates on the large-scale photometric and morphological
properties for the galaxies in our sample.
We will first describe the observations and data reduction
procedures.  Then, we comment on the UV photometric
properties of the galaxies (e.g., IRX-$\beta$).  We give detailed descriptions
of each system, then discuss their detectability at high-$z$.  Finally, we
consider our systems in relation to Lyman-break galaxies, extremely red 
objects, and sub-mm galaxies.
Here, we assume a Hubble Constant of $H_0=70$ km/s/Mpc, with 
$\Omega_M=0.3$ and $\Omega_\Lambda=0.7$.
Computing angular size and luminosity distances in a cosmology with 
nonzero $\Lambda$ is described by \citet{car92}.

\section{Observations}
We constructed a flux- and luminosity-limited sample of luminous infrared
galaxies (LIGS: $L_{IR}=10^{11}-10^{12}L_{\sun}$)
and ULIGs 
from the IRAS Bright Galaxy Sample, then selected from that list 
seven relatively nearby galaxies that had already been imaged at 
other wavelengths by HST.  
Our sample spans a factor of 4 in IR luminosity, including 2 LIGs
at the highest luminosities of the starbursts studied in the papers by
Meurer et al., $4\times10^{11}L_{\sun}$, and 5 ULIGs,
going up to $1.5\times 10^{12}L_{\sun}$ (listed in order of
increasing right ascension in Table 1).

We obtained our far- and near-UV images using the MAMA detectors 
of the Space Telescope Imaging Spectrograph (STIS)
on HST.  In
the near-UV channel (STIS NUV-MAMA+F25QTZ filter, 
$\lambda_{pivot}=2364\mbox{\AA}$,
$\Delta\lambda=903\mbox{\AA}$\ FWHM), we obtained two 
exposures totalling 1500 
seconds, with the telescope offset by 0\farcs 24 in RA and DEC 
between exposures.

The far-UV observations (STIS FUV-MAMA+F25SRF2 filter, 
$\lambda_{pivot}=1457\mbox{\AA}$, $\Delta\lambda=244\mbox{\AA}$ FWHM), 
were divided into 2-3 similarly offset pairs 
of images, for far-UV exposure times of 2829--3891 seconds.  
Details of the exposures are given in Table 2.

\section{Analysis}
We analyzed the STScI pipeline-processed STIS data 
as follows.  First, we split the paired far-UV images into
single ones, giving typically six far-UV images and two near-UV images
per galaxy.
For each individual image, we determined and subtracted the mean values of the 
sky found from large regions well away from detectable galaxy emission.  
The far-UV background was found to vary significantly in some cases, due 
to light near the limb of the Earth.  Because the extended emission
from our sources is typically 1--2 DN per pixel, a few images 
with high backgrounds ($>1-2$ DN per pixel) 
were rejected, because the S/N ratios of the combined images would have been
limited by the Poisson noise of the high-sky images.
The images were aligned and co-added
using the DRIZZLE package in IRAF (v2.10), with which 
we also corrected the geometric distortion (using the cubic distortion
coefficients from the STIS Instrument Handbook, Version 3, June, 1999)
and rotated the images to the standard
North-up, East-left orientation.  We also magnified the far-UV images
to the same platescale as the near-UV images (0\farcs 024465/pixel) 
during the DRIZZLE process.

For the STIS images, determining accurate sky values was important
because much of the total UV flux comes from large regions with surface
brightnesses near the residual sky levels.  The uncertainty in our photometry 
is therefore quite sensitive to accurate subtraction of the residual sky
levels.  We measured the sky counts in the darkest regions of the
images, in a few contiguous groups of 20$\times$20 pixel boxes, about
30--40 boxes in total for each final UV image.  We determined the
residual sky levels for each of the final images from the averages of
the means of the sky counts in the boxes, and the uncertainties in
the residual sky levels were set to the $1\sigma$ scatter in the means
of the sky counts of the boxes.  One galaxy, Arp 220, was especially
problematic, because its large angular size resulted
in faint emission essentially filling the entire STIS field of view.
For Arp 220, the darkest regions of the UV images turned out to be
located within the dark dust lanes apparent in optical images.\footnote{
For Arp 220, the raw NUV sky values are heavily quantized at around
0 and 1 DN/pixel.  The mean values of the raw NUV skies in large, 
dark areas of the 
images, in these 750 second exposures, were $\sim$0.1 DN/pixel.  
The FUV sky means, in large sections of the raw
UV images, varied from 0.004 DN/pixel to 0.68 DN/pixel from image
to image.  The variation was
due to scattered light from near Earth's limb in the images. So, the 
highest the true FUV (sky+galaxy) value 
could have been was 0.004 DN/pixel, in an exposure of 690 seconds.
Any extended emission from the galaxy can have at most that much flux.
After combining the FUV images with those sky values subtracted,
the final value of the FUV ``residual" sky was $<1\sigma$ from 
zero.  Our error estimates should adequately incorporate 
the possibility of
errors in the sky determination. In short, the extremely low 
counts in dark areas of the raw Arp 220 UV
images lead us to believe that we are not losing significant amounts of
uniform, extended light across the images.}
In all other galaxies, comparison with WFPC2 (or NICMOS) images show
that the STIS field of view was large enough that we should have 
regions of blank sky in our STIS images, and those regions with lowest counts
were used to determine the sky levels.

Measuring total fluxes is
difficult because the objects have irregular shapes, and in several cases
their surface brightnesses barely ever reach above the sky level over
much of the face of the galaxy.  We developed a ``buffered
isophote'' technique for measuring total fluxes.  First, we used the
adaptive smoothing algorithm of Scoville et al. (2000) to make a
smoothed FUV image of each galaxy.  This was used to make a ``good pixel"
mask of all pixels in the smoothed image that are $\geq 3\sigma$
above the background, where $\sigma$ is the scatter about the mean
background as described above.  It is important to use the smoothed
image for this so that the fluxes are not biased upwards by positive
noise deviations.  Since the light distribution undoubtedly extends
beyond the $3\sigma$ isophote, we then buffered this mask by appending
all pixels within 1$''$ of the previously marked pixels as also being
``good."  The choice of 1$''$ as the buffer size is somewhat arbitrary,
but by blinking the mask and the smoothed images it was clear that there
is little or no real structure beyond the mask.  The total flux is
then found by adding the flux in the ``good" pixels of \emph{the original
images} (before smoothing).  Both the FUV and NUV fluxes were extracted
with the FUV mask.  Using the NUV image to create the mask often
resulted in slightly larger fluxes, but the differences were within the
photometric errors.  We also performed simple circular aperture
photometry 
in apertures of 0.5, 1, 2, 3, and 5 kpc physical radius in our
assumed cosmology (Table 3).  We note that the larger
apertures can include flux from
companion or field galaxies in some cases.  Errors were determined from
photon statistics 
and the uncertainties in the sky levels (the latter
dominates in all cases).  In all but one case (IRAS 19254--7245), at
least 85\% of the ``total" flux is contained within the 5 kpc radius
aperture (or the sum of the two apertures for IRAS 08572+3915 and VV 114).
We note that photometry of IRAS 19254--7245 in large apertures is 
intrinsically uncertain at the
level of 0.26--0.46 magnitudes because of the very low flux levels, and
some of the ``missed" flux likely comes from a few star clusters outside
the 5 kpc apertures.

WFPC2 images of the five of our galaxies with $L_{IR}>10^{12}L_{\sun}$ 
had been taken, and we retrieved
those from the HST Archive.  We co-added the pipeline-calibrated 
images, using CRREJ and COSMICRAYS to correct for
cosmic rays and hot 
pixels.  We then used DRIZZLE to perform the 
geometric corrections, adjust the pixel scale to that of the STIS images, 
and rotate the images to the standard North-up 
orientation. 

NICMOS images of all but one of our galaxies (IRAS 19254--7245) were taken 
and made available in 
reduced form by \citet{sco00}.  We used MAGNIFY and ROTATE to set the pixel
scale and orientation to those of the STIS images.
In all cases,
there were enough common details between 
the UV and optical, and near-IR and optical, that we could confidently 
co-register the multi-wavelength data relative to the WFPC2 
images using IMSHIFT.  (For the two galaxies without WFPC2 images, 
VV 114 and IC 883, we
aligned the UV and IR images relative to the F110W and F160W images.)   
We estimate registration uncertainties 
of $\la$ 1 STIS pixel.

\section{Results}
\subsection{Global UV properties} 
\subsubsection{The relationship between UV color 
and the FIR/UV excess}
As noted in Section 1, \citet{meu99} (hereafter M99)
describe an empirical relationship between the color of
the UV continuum (parameterized by $\beta$) and the FIR/UV flux ratio
for a sample of starbursts observed with IUE -- the IRX-$\beta$
correlation.  The FIR emission results from dust heated by the UV
photons it absorbs.  Hence, for galaxies obeying this correlation, one
can use $\beta$ to predict the FIR/UV flux ratio, and hence recover the 
dust-absorbed
UV emission.  This can be especially useful at high redshift, where
access to the rest-frame FIR is currently very difficult.  One problem
with this method is the heterogeneous nature of the calibrating IUE
galaxy sample.  It is not yet clear how well this calibration works for
FIR-selected (e.g. IRAS) galaxies.  We can now begin to address
this issue for the most luminous IRAS galaxies.

We 
have calculated the global values of $\beta$ and
the IRX for our ULIGs
(see appendix) after first correcting for Galactic extinction using the
values in Table 1.
and for the galaxies studied by TKS using the fluxes in that paper.  The
values for our ULIGs are given in Table 3.  We plot $\beta$ and the IRX in
different apertures in Figs. \ref{global_betairx} and \ref{nuclear_betairx}.

The key result is that most of our LIGs and ULIGs
do not obey the IRX-$\beta$ correlation, irrespective of aperture size.
The sense of
this is that in most cases, our galaxies have much higher IR-excesses
for their UV colors than M99's galaxies.  In the absence of the
FIR data, using the M99 correlation would lead us to greatly
underestimate the FIR fluxes of
most of our galaxies. 
One might have hoped that if the bluer light from star-forming regions far
from the nuclei were ignored, the light from the nuclei themselves might
follow the IRX-$\beta$ correlation, but this is not the case.
The discrepancy between our galaxies and the M99 sample
gets much worse as we decrease the aperture sizes
and split the FIR fluxes between members of interacting pairs 
(Fig. \ref{nuclear_betairx}).
By far the worse failure
is Arp 220, where using the 2 kpc diameter aperture would have us 
underestimating the FIR flux by a factor of $\sim 10^4$.
While the UV colors are indeed redder in the smaller apertures, the UV
fluxes themselves drop off as well.
On the other hand, VV 114,
the second least luminous system in our sample, comes the closest to
obeying the IRX-$\beta$ correlation, falling within the scatter of IUE
starbursts.  VV 114 is composed of UV-bright and UV-faint components,
and it is the former that is similar to M99's starbursts.  (It is possible
that the nucleus of IRAS 19254-7245 S is consistent with IRX-$\beta$, 
since we only have a lower limit for $\beta$.)

One possible interpretation of the above two results is that 
the IR-luminous regions of the galaxies are so obscured that we have
detected no UV light from them at all.  In this case, the UV light we do
see comes from relatively lightly-reddened star-forming regions and 
stars outside
the most obscured regions, giving blue colors.  
We would then
expect the IRX to be essentially uncorrelated with color, as
is observed.  Inspection of Table 3 shows that only on order of a few
percent of the total NUV flux arises from the inner 0.5 kpc radius
(from 0.07\% for VV 114 E to 7.3\% for IRAS 15250+3609).  
This is 
also consistent with the central, intrinsically luminous regions being
hidden from view.
 This is the expectation of radiative transfer
models where stars and dust are mixed, such as the ``dusty'' model with
SMC type dust of \citet{wit00}; see their
Fig.~12\footnote{which is mislabeled Fig. 13 in the print version}.
However, for a given geometry, the Witt \& Gordon models cannot explain
both the range of the IRX-$\beta$ correlation seen in M99's starbursts
(which seems to require
foreground dust) and the position of the LIGs and ULIGs (which does not).
Either the relative distribution of dust and stars is different between
the IUE starbursts and our sample, or a more complicated dust geometry
than that explored by Witt \&\ Gordon is required.

This picture has important implications if ULIGs
are in fact the analogues of the sub-mm galaxies.
We should not expect optical spectroscopy of the
sub-mm sources, probing the rest-frame vacuum UV, to be sensitive
to the dominant luminosity source in the inner few hundred pc 
of the galaxies.  In some cases, an AGN may
be seen through a hole in the obscuring gas and dust; but the absence of
an AGN spectrum is not sufficient to rule out the possibility of a buried
AGN.

\subsubsection{Optical depths in the ULIGs}
Sub-mm line and continuum observations of ULIGs have shown that
they contain tremendous concentrations of gas and dust in their nuclei
\citep{sco91,san91,dow98,bry99}.  Commonly, $10^{10} M_{\odot}$ 
of gas are located
in the inner kpc of ULIGs.  The highest spatial resolution
observations show the existence of what \citet{dow98} called
``extreme starburst regions," wherein $\sim10^{9} M_{\odot}$
of H$_2$ gas is concentrated in regions typically 100 pc in size.
The resulting beam-averaged surface densities of dust (assuming
reasonable gas/dust ratios and spherical symmetry) are so high 
as to give extinctions of up to $10^2-10^3$ 
magnitudes $A_V$ in the inner few hundred pc.  

This simple model, which requires $A_K \approx 0.1 A_V \approx 10-100$ 
magnitudes,
cannot apply to all parts of the nuclear regions in the ULIGs, since bright 
near-IR emission is in fact
seen from the nuclear regions of all ULIGs (e.g., Scoville et al. 2000).
Further, we see UV emission concentrated towards the inner few 
hundred pc in some of the
galaxies, even though the compact nuclear sources are usually not well
detected in our images.  An alternate picture, as suggested by \citet{dow98},
might be a compact,
totally obscured (in the UV, near-IR, and even mid-IR) core 
(the extreme starburst regions) with 
still dense, yet not entirely opaque (in the near-IR) 
mixed gas/dust clouds 
on the exterior (the surrounding ``molecular disk").  On the outside would
be additional mixed gas/dust clouds and UV sources, possibly corresponding
to ``field" star formation.  This would 
allow us to see some relatively blue
extended emission and star clusters outside the most obscured regions.
If this is true, some of the UV light visible at the apparent IR nuclei would
come from foreground stars seen in projection.

\subsection{Implications for Lyman-break galaxies}
\citet{ade00} suggested that LBGs and ULIGs/sub-mm galaxies are the extreme
objects along an increasingly reddened sequence of starbursts.  Their paper
marks the most detailed attempt yet to unify these apparently disparate
objects.

Adelberger \& Steidel examined the SEDs of a sample of ULIGs in the mid-IR, 
FIR, sub-mm, and
radio.  After computing an ``average SED," they showed that one could predict
the bolometric luminosity of a ULIG from observations at any one of those
wavelengths, with typical uncertainties of $\sim$0.2-0.3 dex.  
\footnote{This relationship is empirical; it can be explained qualitatively 
by the fact that the dust temperatures, calculated from the 
60/100 \micron\ flux ratios, span only a factor of $\sim 2$ among ULIGs; 
the sub-mm observations are in the Rayleigh-Jeans regime; and the
radio-FIR correlation has a small dispersion.}

They then went further, showing that the admittedly few radio and mid-IR 
observations of $z\sim 1$ galaxies are
consistent with the IRX-$\beta$ relation found for the
M99 sample of local starbursts.  
The existence of IR-luminous galaxies at $z\sim 1$ has been inferred
from {\em Infrared Space Observatory} observations, particularly of the
Hubble Deep Field \citep{aus99}.
About 16 galaxies with $z\sim 1$ were detected in the 15 \micron\ images.
At these redshifts, the ISO observations are sensitive to 
the broad, rest-frame polycyclic aromatic hydrocarbon (PAH) emission features
seen in local starbursts.  Adelberger \& Steidel showed that the 
FIR luminosity of
local ULIGs can be estimated to an accuracy of
about 0.22 dex from the strength of these mid-IR features.  If that
relationship holds at $z\sim 1$, then the ISO-detected galaxies would
have FIR luminosities like local ULIGs, but would have lower values of
the IRX, 
approximately following M99's IRX-$\beta$ relation.  This is 
some suggestion that
while ULIGs exist at $z\sim 1$, they might be different (more
transparent) than local ULIGs.

The IRX-$\beta$ relation might hold at $z>1$ as well.
Rest-frame UV observations
of two high-$z$ galaxies, ERO J164502+4626.4 and SMM J14011+0252,
are consistent, within errors, with those galaxies obeying the IRX-$\beta$
relationship.  Finally, measurements of the ensemble fluxes (at 850 \micron\
and radio wavelengths) of samples of $z\sim 3$ LBGs show that these galaxies 
are at least not grossly inconsistent with IRX-$\beta$.  The overall 
conclusion of Adelberger \& Steidel is that observations at $z\approx(0,1,3)$
are all consistent with the existence of an IRX-$\beta$-like relationship
holding all the way to high redshifts.  In addition, computing the total 
$L_{tot}\sim(L_{UV}+L_{dust})$, Adelberger \& Steidel show that galaxies
with higher $L_{tot}$ are more obscured (that is, have a higher value
of the IRX) than less bolometrically luminous galaxies.  However, it
also appears that the slope or zero-point of this relationship evolves
with redshift, in that for a given star formation rate, galaxies are
more transparent at $z\approx 3$ than at $z\approx 0$.  

To further complicate matters, there are some LBGs with fairly high
luminosities ($\sim 10^{11}L_{\odot}$) that appear to have very little
dust (as determined by the absence of appreciable dust reddening).  These
objects will have relatively low FIR luminosities if they obey the 
IRX-$\beta$ correlation.  The LBGs expected to have the
highest bolometric luminosities have similar apparent UV magnitudes, but
are much redder.  They have predicted $L_{bol} \sim 10^{12} L_{\odot}$ 
(M99), while the brightest sub-mm galaxies are estimated to be ten
times more luminous.
Adelberger \& Steidel attempted to logically connect these lower-luminosity 
objects
that are detectable in deep optical surveys, to the higher-luminosity 
dusty objects (sub-mm galaxies are the most extreme examples) 
that were present in the early universe,
yet are not easily detected (if at all) with the same techniques.

As noted earlier, the IRX-$\beta$ relation at $z\approx 0$ was 
determined by M99 from observations
of a heterogeneous sample of fairly UV-bright starbursts.  That sample
was not preselected to contain galaxies with high $L_{IR}$, so it was not
clear that LIGs and ULIGs obeyed the IRX-$\beta$ relationship.  The
HST observations of three ULIGs by TKS were the first indication that
ULIGs in fact deviated from IRX-$\beta$.  Our new observations confirm
that clearly, for the ULIGs in our sample, IRX-$\beta$ does not hold.  
The IRX-$\beta$ relation appears to break down at 
${\rm log}\left(\frac{F_{FIR}}{F_{FUV}}\right)\sim 2$.  
Though \citet{ade00} showed that two sub-mm galaxies were consistent
with IRX-$\beta$, this would not, in general, be the case if our ULIGs
are good analogues to sub-mm galaxies.

One point, however, is that
not all our galaxies fail IRX-$\beta$ completely: the LIG VV 114 in
particular is not too far off.  There are two galaxies in the M99 sample
with $L_{IR}\approx 4\times 10^{11}L_{\odot}$. One,
NGC 3256, obeys IRX-$\beta$, while the other, NGC 1614, 
is overbright in the FIR.  It would be interesting to observe 
more galaxies
with $L_{IR}=10^{11}-10^{12} L_{\odot}$ to see if there is some
particular value of $L_{IR}$ where galaxies begin to move far off the
IRX-$\beta$ relationship.  Our sample is too limited to tell, as 
it contains only two galaxies in that luminosity range.

We have referred to the breakdown of IRX-$\beta$ primarily as a
luminosity effect, but it may be that dust content and spatial distribution
are more important.  Dust content and FIR luminosity are correlated,
so compared to low-luminosity starbursts, ULIGs have greater amounts of 
gas and dust 
available to both fuel and obscure the compact starbursts (or AGN) 
that provide the high luminosities.  Additionally, in ULIGs, this gas and
dust is centrally concentrated, usually in the inner kpc or less,
resulting in immense optical depths.

\subsection{Starbursts vs. AGN}
Many of our galaxies evidence strong, nuclear activity not necessarily
connected to star formation.  Detailed descriptions of each galaxy
are given below, but
of the galaxies in our sample, Mrk 273 and the southern 
(IR-dominant) component of IRAS 19254--7245 have 
Seyfert 2-type optical spectra; the near- and mid-IR properties of
VV 114 give evidence that an AGN might be present in the eastern
half of the system; and IRAS 08572+3915 NW, IC 883, IRAS 15250+3609,
and Arp 220 have LINER spectra.  The connection between LINER spectra 
and AGN activity
is still poorly understood; a LINER spectrum might be due to shock-excited
gas associated with a starburst-induced ``superwind," and have no
relationship to (black hole-inspired) 
AGN activity at all \citep{ho99,bar00,lut99}.

The best evidence for ``dominant" AGN (i.e., producing $\geq 50\%$ of
the FIR flux) is probably in IRAS 08572+3915 NW and IRAS 19254-7245 S.  
On the other hand, the presence of strong starburst activity (perhaps
with a modest AGN contribution) seems 
well established in the other five galaxies in our sample.  This is
detailed below in \S4.4.

How would the presence of an AGN affect the IRX-$\beta$
correlation?  The starburst fraction of the FIR emission would decrease
as the AGN fraction increased. If we detect only the starburst in
the UV, and if only the starburst fraction of the FIR emission were
counted, this would push the galaxy downward on the
IRX-$\beta$ plots, closer to the relationship found for the
UV-selected starburst sample.  As noted above, AGN are likely to make
a substantial contribution to the FIR in only two cases.  VV 114 falls
on the IRX-$\beta$ relationship, the other four likely to 
be starburst-dominated all fall far above.
Thus, a minor AGN contribution to the FIR is not capable of
explaining the weak UV detections.

\subsection{Individual galaxies}
\subsubsection{VV 114}
VV 114 (IC1623, Arp 236) was discovered to be a bright FIR source in the
original IRAS Bright Galaxy Survey.   At about 87 Mpc distance, it is
one of the more nearby galaxies with such a high FIR luminosity.  

Optical and near-IR imaging \citep{kno94, doy95} show a strong contrast
between the two members of the merging pair that comprise VV114.  The
western component (VV 114 W) is dominated by a blue, high surface brightness
complex of regions with a relatively weak near-IR nucleus.  The
eastern component (VV 114 E) is much redder and brighter in the
near-IR, and contains two prominent, extended sources: A and B of Doyon
et al.  Only the brighter of these, source A, is apparent at 3.8\micron.
The HST NICMOS images of \citet{sco00} show that source A contains a
bright, pointlike source that had been unresolved in the ground-based
images.  

Optical and near-IR spectra indicate that star formation dominates the
energetics of VV 114 W, while both star formation and an AGN may
power VV114E.  At 2.2\micron, the latter shows a very red continuum
with starburst features (He{\sc i} and Br$\gamma$ nebular emission lines and
stellar CO absorption bands) superimposed
\citep{doy95}.  The near-IR spectrum of region A, can be modelled by a
combination of about 50\% highly reddened starlight ($A_K\sim 0.5$ mag,
for $A_V\sim 5$ mag), and about 50\% hot dust ($T\ga 500K$) which is
consistent with the broad-band near-IR colors.  The high hot dust fraction
of the concentrated red emission in source A is very
unusual in what otherwise seems to be a starburst
\citep{gol95,gol97a,gol97b}. The 5--12 \micron\
colors (from the Infrared Space Observatory, ISO) are also consistent 
with strong
hot dust emission, leading \citet{lau00a} to suggest that VV 114 E may
get up to one-half of its mid-IR luminosity from an AGN.  Maps of VV 114
E at 8.44 GHz \citep{con91} show the multiple morphology seen at 2.2
\micron, and the radio counterpart of source A contains several
small sources.  At lower resolution, radio maps \citep{con90} show that
21 cm continuum radiation is widespread throughout the system with
concentrations corresponding to both VV114W and E. Sub-mm line and
continuum maps \citep{yun94, fra99} show great amounts of molecular gas
and dust are present in VV 114, with morphology grossly similar to the 21 cm
maps.  Much of the material detected in the sub-mm maps is located
between VV 114 E and VV 114 W.

The two components of VV114 are probably nearly equal in total
luminosity, although their overall SEDs are very different.  Using the
21 cm flux as a tracer of FIR, \citet{kno94} suggest that the FIR flux
is split 60\% from E, 40\% from W.  However, with this split, 
$F_{FUV}/F_{FIR} \sim 0.2$ for the W component, which means that the FUV
flux also contributes significantly to the bolometric luminosity.  From
Starburst99 models \citep{lei99} we estimate a bolometric correction of
$\sim 1.3$ for the STIS FUV fluxes of star forming populations
\footnote{This refers specificially to a 100 Myr duration
constant star formation rate population having solar metallicity, and a
Salpeter IMF with upper mass limit of 100 $M_{\odot}$.}, hence the
detected FUV emission adds another $\sim 27$\% to the bolometric flux of
the W component.  Therefore in terms of bolometric luminosity, the split
is about 45\%\ in the E and 55\%\ in the W components.

The contrast betwen the two components is most extreme in our UV
images (Figs. \ref{fig_vv114a}-\ref{fig_vv114c}).  These show exquisite detail, 
and the greatest number of star
clusters observed in any of our galaxies.  Several hundred clusters 
are seen
in VV 114 W, many visible all the way from the far-UV through near-IR.
However, very little UV emission is detected from VV 114 E.  No bright UV
sources are seen in the near-IR bright regions of VV 114 E.  No
detectable UV sources are seen projected within $\sim 250$ pc of B,
while two faint, fairly blue UV sources are seen projected about 100 pc
from source A.  The fairly blue colors of the clusters in VV 114 W, and
the similarity of the UV and near-IR images of VV 114 W, suggest that
dust absorption is not nearly as strong there as in VV 114 E.

\subsubsection{IRAS 08572+3915}
IRAS 08572+3915 consists of an obviously interacting, yet still
well-separated pair of galaxies.  All available data point to the NW
component (``IRAS 08572+3915 NW") as being responsible for nearly all of
the far-IR luminosity of the pair.  It is the only one of the pair
detected in mid-IR and radio maps.  The other galaxy (``IRAS 08572+3915
SE") has optical colors similar to quiescent galaxy nuclei and is not
detected in VLA maps \citep{con90} (inspection of those maps shows that
the peak surface brightness at the position of SE is $\leq 10$\% of that
at NW) or in the mid-IR, where Soifer et al. 2000 limit the SE
source's contribution to $\la 30$\% of the total mid-IR flux).
Using these non-detections as a guide, we
conservatively assign 30\%\ of $F_{\rm FIR}$ to SE (an upper limit, really), 
and 70\%\ to NW.

There is good evidence that the nucleus of IRAS 08572+3915 NW is very
dusty, possibly harboring an AGN.  In optical WFPC2 images
\citep{sur98}, the nucleus is not apparent in the blue ({\em F439W}) and
only begins to show in the $I$ band ({\em F814W}), while the optical
spectrum is that of a LINER \citep{vei95}.  In the near-IR it is
extraordinarily red at 2.2 \micron, $f_{\lambda}\propto \lambda^3$, the
reddest of the 56 nuclei studied in the $K$ band spectroscopic study
of \citet{gol95, gol97a}.  In the mid-IR, the morphology is compact, the
colors are warm, and the spectrum shows a deep silicate absorption
feature \citep{soi00, dud97, deg85} all of which are indicative of the
presence of an AGN.

Our UV images (Figs. \ref{fig_ir08572a}-\ref{fig_ir08572c}) 
reveal a morphology quite similar to that seen in the HST
{\em F439W} images of \citet{sur98}.  Bright emission is seen from the
nucleus of IRAS 08572+3915 SE, and H{\sc II} regions or star clusters
are seen in its southern spiral arm; diffuse emission is seen throughout
that entire galaxy.  Diffuse emission is also seen in the disk of IRAS
08572+3915 NW.  However, no peak in the UV emission is seen at the
small, red {\em F814W} source identified as the nucleus.

\subsubsection{IC 883}
IC 883 (Arp 193, UGC 8387)
shows the appearance of a single, edge-on
galaxy, oriented roughly SE-NW.  Dust lanes cross through the nuclear 
region, and reddening may be significant even in the near-IR. 
A number of objects, probably star clusters, are seen on either side 
of the disk \citep{sco00}.

Optical spectroscopy \citep{vei95} shows IC 883 to have a LINER spectrum.
Spectroscopy at 2.2 \micron\ \citep{gol97a} reveals starburst 
emission lines and
stellar CO absorption features.  The relatively deep CO index and large
Br$\gamma$ equivalent width argue that the great majority of the continuum
at 2.2 \micron\ is of stellar origin.

The nuclear region shows two peaks about 0\farcs75 apart, embedded 
in extended emission, at
wavelengths ranging from the near-IR, to the mid-IR, and to the radio
\citep{sco00,soi01,con91}.  However, the two peaks 
do not appear to overlap precisely; their position angles and separations
are different in the radio and at 2.2 \micron.
This might be due to high obscuration even at 
2.2 \micron\, or it could be due to multiple sources in the inner 
$\sim$1.5\arcsec, with 
great differences in colors.  Given the uncertainties in the HST astrometry, 
we are unable precisely align the radio and HST images,
and for the remainder of this paper, we assume that the brightest
source at 2.2 \micron\ is coincident with the brightest source at
8.44 GHz.  This alignment results a good match of the kpc-scale radio 
and 2.2 \micron\ light. \footnote{\citet{soi01} 
assumed that the same two peaks are visible in both the radio
and IR images, accepting the separation and position angle differences.
They noted that their alignment resulted in a poor
match between the near-IR and radio kpc-scale emission.
While their alignment may well be shown to be correct
in the end, it does not substantively affect our conclusions.}
Sub-mm observations show an elongated morphology,
but do not resolve any pointlike sources \citep{dow98}.

Our UV images (Figs. \ref{fig_ic883a}-\ref{fig_ic883b}) 
show faint, patchy extended emission. 
The extended emission is at a minimum where the
near-IR emission is strongest, and there are no prominent UV peaks
associated with the nuclear near-IR peaks.  
There are a few bright pointlike sources in the UV images, 
probably star clusters, most of which have near-IR counterparts.
However, most of the near-IR clusters do not have UV counterparts.
We conclude that the extended region in the nucleus, so bright 
in the near-IR, is quite obscured in the UV.  Unfortunately, our STIS
images are not optimally centered, and the galaxy probably extends outside
our images.

\subsubsection{Markarian 273}
Markarian 273 is a visually striking merging system, with an amorphous 
central region, a tidal tail stretching $\sim$40 kpc nearly 
due south of the center, and a ``tidal loop" counter-tail to 
the north.  Sub-arcsecond resolution imaging from the optical to
the radio has revealed three primary sources 
within the inner 2\arcsec\ of the galaxy:
Mrk 273 N, Mrk 273 SW, and Mrk 273 SE.  
\citep{kna97, soi00, con91}.

Both starburst and AGN activity are found in the central kpc. 
Mrk 273 has a reddened Seyfert 2 optical emission line spectrum \citep{vei99}. 
Yet, a spectrum extending more to the blue (covering
3400-5500$\mbox{\AA}$) presented by \citet{gon01} is dominated by a reddened
stellar continuum with strong Balmer absorption lines.  In fact, this
new spectrum is well matched by a combination of young, intermediate age, and 
old stellar populations, though potentially up to $\sim$20\% of the 
optical light could come from a power-law (AGN?) continuum.  While the
near-IR spectrum is that of a starburst \citep{gol95}, high-excitation lines
signalling an AGN are seen in the mid-IR \citep{gen98}.
X-ray observations from the ASCA \citep{iwa99} and BeppoSAX \citep{ris00}
satellites indicated the presence of an obscured, possibly 
variable, hard X-ray source in the nuclear region.  Higher spatial resolution
observations from 
Chandra \citep{xia01} have now revealed a compact, hard X-ray
source positionally coincident with Mrk 273 N.  Diffuse, softer 
X-rays are seen from the surrounding nuclear region, and 
in fact envelop the 
entire optical structure of the galaxy.  No X-ray point sources are
found at either Mrk 273 SE or Mrk 273 SW.
Unfortunately and importantly, the fraction of $L_{\rm bol}$ produced 
by the AGN is still not known.

Mrk 273 N is clearly detected as bright,
extended region in the near-IR and mid-IR, and some faint 
optical emission is seen within 0\farcs 1 of the near-IR peak.  
Very high-resolution radio observations show that Mrk 273 N is 
itself bifurcated at the 0\farcs 1 scale, and may be a rotating disk
\citep{car00}.  
Mrk 273 N appears to be a galactic 
nucleus---but whether it is the merged nucleus, or the 
nucleus of only one of the merging galaxies, is not clear.  

A strong radio source, Mrk 273 SE is located 0\farcs 75 from 
Mrk 273 N.  Mrk 273 SE is highly absorbed 
in H{\sc i}, suggesting significant optical obscuration, yet at 
optical and near-IR wavelengths, a blue, rather faint point-like 
source is visible at the radio position \citep{sco00}.  The third major 
emission region, Mrk 273 SW, shows a bright, point-like 
peak in near-IR/mid-IR observations, yet its likely radio counterpart is 
noticeably
offset, and weak relative to Mrk 273 N and Mrk 273 SE.  Several studies have 
speculated about the natures of Mrk 273 SE and Mrk 273 SW, and their
relationship to Mrk 273 N,
but no definitive answers have emerged.  \citet{soi00} speculate that
consideration of both their spectrophotometric images and the spectroscopy
of \citet{dud99} suggests that Mrk 273 N shows silicate absorption at
10 \micron\ (indicative of a deeply buried, luminous source, and consistent
with the possible X-ray AGN detection), and 
Mrk 273 SW 
shows PAH emission (indicative of star formation), but this needs 
to be confirmed.

Our UV images (Figs. \ref{fig_mrk273a}-\ref{fig_mrk273b}) show faint, diffuse 
emission covering an area 
of $\sim 10\arcsec \times 10\arcsec$, with the northern tidal
tail being prominent.
A few small objects also weakly detected in the WFPC2 image, 
possibly star clusters,
are scattered south of the nuclear 
region.  The most intense UV emission is visible in the vicinity of
Mrk 273 N, but is not centered at its radio position.  Within the central
kpc, the strongest UV emission corresponds with the base of the northern
tidal tail, which comes in from the north and apparently terminates 
$\sim 400$ pc from the nucleus.  
No UV peaks are seen at Mrk 273 SE or Mrk 273 SW.  
No prominent UV emission is spatially 
coincident with the radio peaks of any of the three principal 
sources, though perhaps some is {\em associated} with these sources.  
No direct detection of an AGN is made in the UV.  While it is
possible that some of the UV emission in the nuclear region might 
be reflected or scattered light from the AGN, the extended UV light
seems most
likely due to star formation, reinforcing the conclusions of \citet{gon01}.

\subsubsection{IRAS 15250+3609}
IRAS 15250+3609 is one of the original 10 IRAS ULIGs \citep{san88}.  
However, for reasons that are not clear, it has been
poorly studied.  On large scales, the WFPC2 images show a tidal tail
visible in earlier ground-based images, as well as several other
galaxies of mixed morphological type; we speculate
that perhaps they form a group with IRAS 15250+3609.
The central region is complex, with 
several star clusters and extended emission scattered over an 
area about 4\arcsec\ in
diameter.  The WFPC2 images, however, show that the brightest near-IR
source is only partly visible in the optical image, apparently
obscured by a dust lane.  

IRAS 15250+3609 has a LINER-like optical spectrum
and a fairly typical starburst near-IR spectrum
\citep{vei95,gol97a}.
ISO spectroscopy
reveals strong PAH emission at 7.7 \micron, suggesting that
star formation plays a prominent role in this galaxy \citep{gen98}.  
However, the
upper limits on the high-excitation mid-IR lines are not good enough to
rule out the presence of an AGN.
The center of IRAS 15250+3609 contains a compact, yet resolved 
($\theta \sim 0\farcs 2$) radio source \citep{smi98}, consistent with
a concentrated region of star formation.  The available
evidence points 
towards star formation dominating the energetics of IRAS 15250+3609.

Our STIS images (Figs. \ref{fig_ir15250a}-\ref{fig_ir15250b}) show that the UV 
morphology of the central region is 
similar to the optical morphology.  Most of the optically visible
clusters are UV-bright.  While some faint, extended emission is present
throughout much of the nuclear region, no bright UV source is seen 
at the location of the brightest near-IR source.

\subsubsection{Arp 220}
Though Arp 220 has often been called ``the prototypical ultraluminous infrared
galaxy," progress towards understanding this, the nearest 
ULIG, has occurred slowly, and with many about-turns.

Arp 220 has a highly complex optical morphology, with light extending over
at least 40 kpc.  In the midst of an optically dark dust lane, IR and
radio
images revealed two prominent emission regions about 
1\arcsec (370 pc) apart \citep{gra90,con91}. These
mark what may be the nuclei of the merging
galaxies.  VLBI imaging has shown
the radio-emitting regions to contain numerous milliarcsec-scale
sources, which have been interpreted as radio-luminous supernova 
remnants (Smith et al. 1998).

Optical spectroscopy of Arp 220's nuclear region has revealed a LINER 
spectrum, and strong superwind activity is present
\citep{kim98, hec87}.  
Spectroscopy of the two near-IR peaks
reveals very strong stellar CO absorption features in the $K$-band, 
proving that starlight
dominates the continuum at $2.2\micron$, and there is no
evidence of broad emission lines \citep{gol95}.  ISO mid-IR and FIR 
spectroscopy also
reveals only starburst features (Genzel et al. 1998 and references therein).

Downes \& Solomon (1998) presented a detailed argument showing that there
is sufficient detected starburst activity to power Arp 220, and 
thereby excluding the presence of a powerful AGN.  It is probably fair to say
that there is no direct evidence, at this time, in favor of a luminous
AGN in Arp 220.

Our UV images (Figs. \ref{fig_arp220a}-\ref{fig_arp220b}) reveal that the 
central kpc of Arp 220 emits no concentrated
UV light.
However, extremely faint extended emission is present over nearly the
entire STIS field of view, and the darkest areas of our images are located 
within the dust lanes so prominent in optical images.  A small 
number ($\sim3$) of the star clusters visible
in the {\em F814W} image have clear UV counterparts,
and
a small extended patch visible just east of center in the UV images has
a very weak counterpart in the {\em F814W} image.

\subsubsection{IRAS 19254--7245}
IRAS 19254--7245 is best known as the ``Superantennae" \citep{mir91}.  
With tidal tails 
350 kpc across, it is a merger in progress.  Despite 
the similar optical appearances of the two galaxies, ISO observations
\citep{lau00a, lau00b}
indicate that a Seyfert 2 nucleus \citep{mir91} in the southern 
galaxy produces at least 70\% of the mid-IR luminosity and perhaps
almost all the FIR luminosity.

The {\em F814W} images show that
both nuclei are extended, with complex
morphologies at even 0\farcs1 scales.
We identify the bright {\em F814W} source in the southern 
nucleus as the AGN, since
the optical spectrum is dominated by AGN features.  
Our STIS images (Figs. \ref{fig_ir19254a}-\ref{fig_ir19254b}) show 
very weak and diffuse FUV and NUV emission from
the spiral arms, plus a few faint star clusters.  Only 
in the NUV image is concentrated emission seen near
the optical nuclei.  There is an elongated
NUV source coincident with the southern nucleus {\em F814W} peak, making 
this galaxy the only one of our sample where a reasonably
prominent UV source is clearly
detected at the supposed IR peak.  Though readily apparent, however, 
the source
is not very bright.  In the 0.5 kpc radius aperture, the southern
nucleus is 
almost 2.2 magnitudes fainter in the NUV than
IRAS 15250+3609, and more than 1.7 magnitudes
fainter than IRAS 08572+3915 SE (corrected for 
Milky Way extinction), both at comparable distances.
Further, since the NUV structure of the southern nucleus is clearly extended,
we suggest that it may trace an edge-on circumnuclear disk, 
a jet, or even a star-forming region, rather than the AGN itself.

\section{Discussion}
\subsection{Detecting ULIGs in the UV}
Including the three ULIGs from TKS, all ULIGs observed in the
vacuum UV by HST have been detected.  These galaxies often show both
extended emission from unresolved stars, and luminous star clusters.  In
our sample, the
UV emission tends to be (5/7 cases) strongest projected against the inner
kpc, yet even then it is often faint and there are two cases (VV 114 E
and Arp 220) where very little emission comes from the inner kpc.
However, in 6 of 7 galaxies, there are significant offsets, several 
hundred pc or more,
between the UV peaks and near-IR peaks.
If the near-IR and radio morphologies
of the galaxies in our sample
accurately represent the morphologies of the dominant FIR energy
sources, then our UV images do not, in general, pinpoint the locations of
principal energy generation.

\subsection{Detecting ULIGs at high redshift}
An important part of our study is to determine how these ULIGs would
appear at high redshifts, in deep surveys such as the Hubble Deep Field.

For a galaxy with rest-frame monochromatic luminosity $L_{\nu}$, 
observed at some high redshift with luminosity 
distance $d_{L}$, 
the observed monochromatic
flux density $f_{\nu}$ is
$f_{\nu}=\frac{L_{\nu}\times (1+z)}{4\pi d_L^{2}}$.
The flux density at some high redshift $z_{\rm high}$, in terms of the
flux density observed at the true redshift $z_{\rm true}$ is then:
\begin{equation}
f_{\nu, z_{\rm high}}=f_{\nu, z_{\rm true}}\times 
\frac{d_{L, z_{\rm true}}^{2}(1+z_{\rm high})}
{d_{L, z_{\rm high}}^{2}(1+z_{\rm true})}.
\end{equation}
We neglect the fact that, since our ULIGs are at an average 
redshift of $z=0.046$, we are observing light emitted at slightly shorter
wavelengths than the pivot wavelengths of our UV filters.  

The light seen from each galaxy in the NUV filter would be redshifted
into the
{\em F606W} filter ($\lambda_{\rm pivot}=6002\mbox{\AA}$) at $z=1.54$.
The light seen from each galaxy in the FUV filter would be redshifted
into the
{\em F606W} filter at $z=3.12$.  We computed the high redshift fluxes after
correcting the data for Milky Way UV foreground extinction.
We have predicted the surface brightnesses (in 
AB magnitudes arcsec$^{-2}$) of the light within the 
half-light radii (Fig.~\ref{halflight}) 
for each component of our sample galaxies.  These are given in
Table 4.  For the single galaxies, an aperture radius of
5 kpc includes $\geq $85\% of ``total" flux (see columns 6 and 7 of
Table 3; aperture growth curves are given in Fig. \ref{halflight}).  
We therefore define the half-light radius as that
containing half of the light in the 5 kpc radius aperture.
\footnote{This definition of the half-light radius fails
for VV 114 E, where photometry in the 5 kpc radius 
aperture is dominated by flux spilling over from VV 114 W.  We therefore
ignore VV 114 E for the high-redshift discussions.}

Though the
total fluxes of at least some of the galaxies are within reach of, e.g.,
the HDF, the surface brightnesses are strongly affected by
the cosmological surface brightness dimming.  
The HDF limiting magnitudes 
in $F606W_{AB}$ is 28.21 ($10\sigma$) for point sources \citep{wil96}.
However, \citet{tot00} show that no galaxies were detected
in the HDF with average surface brightnesses less than 27.5 AB mag 
arcsec$^{-2}$
in {\em F606W}.  We worked out the surface 
brightnesses of our galaxies, using the radii containing
half the light in the 5 kpc physical apertures (Table 4).

We also used the data tables of the HDF galaxies \citep{wil96} 
to compute the surface
brightnesses at {\em F606W}, assuming that $r_1$, the ``first moment radius,"
is equivalent to the half-light radius (Fig.\ref{sb_hdf}).  
We would expect to detect all galaxies except Arp 220 in {\em F606W}
at $z=1.54$, but only VV 114 W and IR 15250+3609 would be above the
detection threshold by $z=3.12$.  Perhaps the next brightest couple 
could be seen at the few $\sigma$ level.  If the surface brightness
detection limit can be pushed a magnitude deeper, then all our sources
would be detectable out to $z=1.5$ and about half would be, if placed
at $z=3.1$.

\subsection{ULIGs and Lyman break galaxies}
A major part of our study was to determine how the properties of our
ULIGs, if placed at $z\sim3$, would compare with those of the
Lyman break galaxies discovered at similar redshifts.  

To this end we compare our ULIGs to the 100 Hubble Deep Field U-band
dropouts studied by M99 in Fig. \ref{udropouts_hdf}.  
This shows our $z
= 3.12$ predictions of $V_{606}$ vs. measured $\beta$ for the ULIGs
compared to the observed $V_{606}$ vs. $V_{606}-I_{814}$ color of the
$U$-dropouts.  The $V_{606}-I_{814}$ color is scaled to $\beta$ using
equation 12 of M99. The dotted line shows the selection limits for
$U$-dropouts used by M99.  This figure is analagous to the top panel
of Fig.~5 of M99, without the absorption correction and without
correcting to absolute magnitude (we do present the absolute FUV
magnitudes of our galaxies, corrected for foreground and also for internal
absorption, in Table 5).

We find that most ULIGs would be too faint to have been selected as
$U$-dropouts by M99.  The flux limits used by M99 were adopted to
conform to those used by \citet{mad96}.  While Fig. \ref{sb_hdf}
shows
that if ULIGs were placed at these redshifts, many would have magnitudes
within the cloud of observed HDF {\em detections}, their fluxes are
so weak that they would not make it into reputable catalogs of
reputable $U$-dropouts.  Typically the selection limits (and hence
detection limits) would have to be pushed 2 magnitudes deeper at $z
\sim 3$ to find exact analogs of our ULIGs if they exist.  In addition
to being too faint, three of the ULIGs would be too red to be selected
as $U$-dropouts.

The key result is that, if placed at $z\sim 3$, only 1 of 7 galaxies in our
sample would 
appear as a {\em U}-band dropout in the HDF. The galaxy VV 114 has the UV
color, luminosity, and size typical of LBGs, and hence makes a good local
analog.  But while the UV light may recover half of the bolometric
luminosity after reddening corrections, it still avoids the dominant
source in the near-IR.
Our data show that ULIGs do not overlap with the brighter Lyman break galaxies.
In most ULIGs, the galaxies are 
so UV-faint and UV-extinguished that they could never be
selected as LBGs in the first place.

\subsection{ULIGs and Extremely Red Objects}
It has recently been determined that
some sub-mm galaxies are Extremely Red Objects (EROs: $R-K\ga 6$ in the
Johnson-Cousins system; Dey et al. 1999, Smail et al. 1999, Frayer et al. 
2000).
Would our galaxies be EROs?  

We predicted the {\em R-}band fluxes of our galaxies at $z=1.96$ (where the
NUV filter would sample the rest-frame {\em R} band) 
and at $z=3.80$ (where the FUV filter would
go to {\em R}) as outlined in \S5.2, converting them
from AB magnitudes to the Johnson-Cousins system in this case.

We estimated the observed {\em K}-band fluxes of our galaxies 
at $z=1.96$ and $z=3.80$ using
optical photometry from the literature, in a manner similar to our calculations
above.  We restricted our optical photometry to two papers, for the sake of
uniformity.  \citet{mir91} obtained photometry of the two components of 
IRAS 19254--7245 through standard
{\em BVRI} filters in an 8\arcsec\ diameter aperture, 
nearly matching our
5 kpc radius aperture (4\farcs 245 radius).  For IRAS 08572+3915 NW, Mrk 273,
IR 15250+3609, and Arp 220, we 
used {\em Bgri} photometry from \citet{san88} in 10\arcsec\ diameter
apertures, which match fairly well the dimension of the 5 kpc radius 
apertures given in Table 3.  
We converted the {\em gri} photometry to
the Johnson system using the transformations given by \citet{thu76} and
\citet{wad79}.  
Our predicted values for $R$, $K$, and $R-K$ are given in Table 6.

Of the IR-dominant nuclei in Table 6, 4 of 5 would have $R-K >4$, 
3 of 5 would have $R-K>5$, and 1 of 5 would have $R-K>6$, for at least
one of the two redshifts.  We note that Arp 220 has $R-K=5.95$ at $z=3.80$, and
so within reasonable uncertainties, might appear as an ERO if at
that redshift, giving us 2 out of 5 of our ULIGs that might be selected as
EROs at an appropriate redshift in the range $1.96 \leq z \leq 3.80$.  
\footnote{However, the ``success"
rate depends strongly on how one defines
it; we might say that 5 galaxies at 2 redshifts gives 10
cases, with only 10\% clearly resulting in an ERO.  Our main goal
is to determine if {\em any} of our galaxies might be selected as an ERO in
the redshift range appropriate to sub-mm galaxies and LBGs, and the answer
appears to be ``yes."}

\subsection{ULIGs and Sub-mm galaxies}
ULIGs are the only objects in the local universe ($z<0.1$) with thermal
FIR emission even approaching that of the sub-mm galaxies.  This makes 
ULIGs the most likely local counterparts of sub-mm galaxies---but is
there a direct match?  
In other words, are ULIGs the local (low-$L$, low-$z$) tail of the 
same class of objects we see
as sub-mm galaxies at $z>1$?  Or are local ULIGs and sub-mm galaxies
different kinds of objects, which both happen to have high FIR luminosities?

First, we return to the fact that
the current sample of
sub-mm galaxies are intrinsically a few times more luminous than any local
ULIG.  Even the most luminous of our ULIGs, Arp 220, 
would have an observed 850 \micron\ flux $\la2$ mJy at $z=1-3$, and
so it would not be present in the current lists of
3$\sigma$ or 4$\sigma$ SCUBA sources.  
The uncomfortable conclusion is that our ULIGs, if placed at
$z>1$, would be difficult or impossible
to detect in either optical or sub-mm deep fields.  

One argument that local ULIGs are not like those at high redshift
can be drawn from the work of \citet{ade00} discussed earlier.  At
$z>1$, IR-luminous galaxies appear to be more transparent in the UV
than local ULIGs.  The reasons for this are not obvious.  One 
possibility,
shown directly in the sub-mm galaxy SMM J14011+0252 galaxy (and hinted 
at in others) by \citet{ivi01},
is that the IR-emitting regions in sub-mm galaxies are much more extended
than in local ULIGs.  This would naturally lead to a greater chance for
``holes" in the obscuration from which UV light could escape.

One way to see if ULIGs and sub-mm galaxies are the same kinds of
object is to ask the question: do the rest-frame UV/optical properties of our
ULIGs match those of the sub-mm galaxies?

\citet{ivi00a} divide the observed sub-mm counterparts
into three classes.
Class II galaxies are bright, with obvious optical counterparts.  Class I
galaxies are bright at $K$, but faint in optical: they are bright EROs.
Class 0 galaxies have no obvious counterparts down to $I\sim26$, $K\sim 21$.
\citet{fra00} examine nine sub-mm sources, and suggest that 40\%--70\% 
of the sub-mm population are in very faint or red galaxies; two of these 
($\sim20$\%) are EROs (Class I).  
Our ULIGs and LIGs
exhibit a wide range of rest-frame UV/optical properties, reminiscent
of the sub-mm galaxies.  
We found that some of our ULIGs 
might appear as
EROs at either $z=1.96$ or $z=3.80$. 
But the sub-mm EROs are bright at $K$, and at those redshifts, our ULIGs 
would be faint.  Only one of the five ULIGs in our sample
(IRAS 08572+3915) would have $K<21$
at $z=1.96$; by $z=3.80$, all would have $K>23.$  Even scaling up the 
UV fluxes from our ULIGs by factors of a few (to match the FIR luminosity
differences) would not be sufficient to render them easily observable.
In summary, our ULIGs (at $z=1.96$ and $z=3.80$) 
would all be fainter
and redder than the Class II sources; mostly fainter than the Class I sources;
and so might only be true counterparts of the Class 0 sources.

Where are the local Class II and Class I counterparts?
We speculate on three possibilities.  First,
the number of ULIGs increases strongly with redshift, so perhaps Class I/II
galaxies are simply vanishingly rare at low redshift.  Second,
perhaps differences in properties such as metallicity result in the 
sub-mm galaxies
being more transparent in the rest-frame UV/optical than
the ULIGs.  Third, while local ULIGs occur almost exclusively in recent or
ongoing mergers, perhaps sub-mm galaxies are made in a different way,
with corresponding differences in properties.  The extended nature of
SMM J14011+0252 hints at this tantalizing possibility.  In the local universe,
generating $>10^{12}L_{\odot}$ in a galaxy appears to virtually always require
a collision and subsequent collapse of the ISM to a dense state in the inner
kpc.  Perhaps galaxies were somehow able to generate these energies without
such collapse at high redshifts.  If this is true, then local ULIGs
would likely not be good counterparts of sub-mm galaxies.  Any real study 
of the differences awaits much more data on sub-mm galaxies.  

\subsection{VV 114, IRAS 08572+3915, and SMM J14011+0252 J1/J2}
In two of our systems, VV 114 and IRAS 08572+3915, we have merging galaxies
where one member of each pair is bright in the UV, but the other is much 
brighter in the near-IR and mid-IR, and (possibly, for VV 114) 
the FIR as well.
If these two systems were sub-mm sources at high redshifts, then optical
observations, sensitive to the rest-frame UV, might not even detect
the less UV-luminous---and more IR-luminous---member of each pair.
The member of the pair responsible for most of the FIR luminosity would
become apparent only in mid-IR observations, which are difficult
to perform from the ground.  On the other hand, optimistically speaking,
we might say that we have {\em two} chances to detect early-stage
merging systems.

\citet{ivi00b} and \citet{sma00} describe the sub-mm selected galaxy
SMM J14011+0252 ($z=2.55$), which may be a high redshift example of this.
That system contains two principal components, the morphologically complex
J1, and the relatively compact J2.  J2 is slightly brighter than J1 at
rest-frame UV wavelength of 1000$\mbox{\AA}$\ (observed {\em U}-band),
whereas J1 is brighter than J2 longward of the
Lyman break (observed {\em B}-band, $\lambda_{rest} \approx 1240\mbox{\AA}$) and
in the rest-frame red optical
(observed {\em K}-band, $\lambda_{rest}\approx 6200\mbox{\AA}$).  
From the above, we know that J1 would be stronger than J2 at the 
rest-frame wavelengths of 
both our FUV and NUV observations.

Very recently, \citet{ivi01} showed that SMM J14011+0252 is significantly
extended in both interferometric CO and radio maps ($\ga 10$ kpc in
the source plane).  Further, the
CO/radio peaks are apparently coincident, but not located at either J1 or J2.
Instead, the CO/radio peak seems to be associated with a newly-discovered,
diffuse ERO component (J1n), about an arc-second north of J1.  This is
reminiscent again of VV 114, where the CO and sub-mm emission both peak
away from the prominent UV/optical/near-IR sources \citet{fra99}.

\citet{ade00} note that SMM J14011+0252 is globally consistent with
the IRX-$\beta$ correlation, though the rest-frame UV flux is the
sum from J1 and J2.  (Likewise, VV 114 is {\em globally} consistent
within the scatter of IRX-$\beta$.)  
Our observations---via analogy to VV 114 and
IRAS 08572+3915---also suggest that J2 need not dominate
the FIR emission from the entire source.  That the radio peak is
apparently associated with a newly-discovered, extended ERO region further
suggests this.
Our study shows that while rest-frame UV images may be useful to
detect merging galaxies, they do a rather poor job of pinpointing
the dominant energy sources, which may be inconspicuous at rest-frame
UV wavelengths.

\section{Conclusions}
We have presented new UV images of a sample of seven LIGs and 
ULIGs taken with the HST.  Our principal results are as follows.

\begin{enumerate}
\item{All seven of our ULIGs were detected in the UV.  They show star
clusters and extended emission, with the brightest UV light
projected within the central kpc of the most IR-luminous members of the systems 
in 5/7 cases.}

\item{However, the UV peaks are displaced 
by at least few hundred pc from the peaks in {\em I}-band and 
near-IR images, and
so presumably the far-IR peaks as well.  At most a few percent of the total
UV light comes from the inner 500 pc, where the majority of the far-IR energy
is generated.}

\item{Most nuclei are reddened.  However, even after correction
for dust reddening using the IRX-$\beta$ correlation, the observed 
light even in large apertures
is insufficient, by factors of 3--75,
to account for the far-IR emission.
When the UV light in 2 kpc diameter apertures is compared with the FIR 
emission, the deficits typically range from 2-4 orders of magnitude;
hence, the reddening is insufficient to account for the large 
infrared-excesses.
We conclude that the compact nuclear starbursts or AGN that dominate
the bolometric energies of these galaxies are highly obscured in the UV.}

\item{All of the galaxies in this study have been previously noted as showing
signs of some AGN activity, usually via Seyfert 2 or LINER spectra.  
However, AGN are not clearly the source of the high infrared excesses in
most cases.  In only two cases, IRAS 08572+3915 NW and IRAS 19254--7245 S,
is it likely that AGN could dominate the bolometric energy output of the
system; the majority appear to be dominated by star formation.  In only
one of the two possible AGN,
IRAS 19254--7245 S, we may detect direct UV emission from the AGN.
Even in this case, there is no emission in the FUV, and the NUV source 
is extended, suggesting that
a circumnuclear disk, jet, or even star formation may be what is
seen in the NUV.}

\item{Artifically redshifting the UV fluxes
shows that six of our galaxies would be detectable in HDF-type
exposures out to redshifts of at least $z=1.54$, yet cosmological surface
brightness dimming would render all but two undetectable by $z=3.12$.  
The galaxies would also have sub-mm fluxes below current
detection limits, if placed in that redshift range. Current optical 
and sub-mm surveys
would therefore probably not detect most of our ULIGs, some of the most 
luminous objects in the local universe, if they were placed at 
redshifts much greater than 1.5.} 

\item{We have compared our galaxies with 100 {\em U}-band 
dropouts at $z\approx 3$ from the HDF.  If placed at $z\sim 3$, most
of our galaxies would be 2--3 magnitudes too faint for robust
selection as {\em U}-band dropouts in catalogs with HDF-like depths.
Only one of our galaxies, the UV-bright member of VV 114,
would have made it into the HDF sample.  This suggests
that ULIGs will not be found in samples of bright Lyman-break galaxies, 
though some could perhaps be lurking amongst the fainter and redder LBGs.}

\item{From estimating the $R$ and {\em K}-band fluxes at
$z=1.96$ and $z=3.80$ for the IR-dominant nuclei of our five ULIGs, we 
find that 
2/5 of them would have colors very close to those of 
Extremely Red Objects.  While some ULIGs may appear as EROs at high-$z$,
in general, their faint but significant UV emission makes them somewhat
too blue to be classified as EROs.}

\item{While our sample is small, the predicted faintnesses and red colors
of our ULIGs, if placed at high redshift, are reminiscent of the
extremely faint/red counterparts of many sub-mm galaxies.  However,
many sub-mm galaxies have counterparts brighter and/or bluer than our
ULIGs would be at $z\approx2$ or $z\approx 3.8$.  One sub-mm
galaxy has been shown to have an extended radio source; if this
is generally true, then that would call
into question the use of ULIGs as low-redshift counterparts.}
\end{enumerate}

\acknowledgments
We thank Neil Trentham for many fruitful discussions concerning
cosmology and his earlier papers, and for reviewing drafts of this
paper.  Nick Scoville kindly provided the
adaptive smoothing algorithm we used, and made his team's NICMOS data
available to the community in reduced form.  
We thank the anonymous referee for insightful comments and suggestions.
Support for Proposal number GO-08201
was provided by NASA through a grant from the Space Telescope Science
Institute, which is operated by the Association of Universities for
Research in Astronomy, Incorporated, under NASA contract NAS5-26555.
This research has made use of the NASA/IPAC Extragalactic Database 
(NED) which is operated by the Jet Propulsion Laboratory, California 
Institute of Technology, under contract with the National 
Aeronautics and Space Administration.  

\appendix
\section{Modeling HST/STIS Fluxes}
The modeling of HST/STIS fluxes was necessary in order to correct the
Galactic foreground extinction within the broad NUV and FUV bandpasses
($\Delta\lambda/\lambda=0.38, 0.17$ respectively) and to calibrate the
relationship between the STIS photometric UV color and the UV spectral
slope $\beta$ ($f_{\lambda}\propto \lambda^\beta$). We employed observed
spectra and continuous and instantaneous star formation synthetic
spectra obtained from the Starburst99 library \citep{lei99}
for a Salpeter IMF slope $\alpha=2.35$ with an upper mass limit of
$100$\,M$_\sun$, metallicities of $Z=0.04,~ 0.02~ and~ 0.004$, and $21$
burst ages ranging from 1 to 900\,Myr.  Fluxes in broad band filters
where measured from the modelled and observed spectra.

\subsection{Galactic Extinction}

We examined the effect of Galactic extinction on the above model spectra
after also applying intrinsic reddening.  This was modelled with the
Calzetti (1997) mean starburst atttenuation curve for $0 \le
E(B-V)_{stars} \le 1$ with increments of $\Delta
E(B-V)_{stars}=0.1$. Each was redshifted to $z=0.04$ -- typical of our
FIR galaxy sample. Finally, the redshifted starburst spectra were
reddened for Galactic foreground using the Galactic extinction curve of
\citet{car89} for $0\le E(B-V)_{Gal}\le1$ with increments of
$\Delta E(B-V)_{Gal}=0.05$. Thus, $220$ combinations of intrinsic and
Galactic reddening were applied to each burst age (21) and metallicity
(3) resulting in $13,860$ model spectra.

The observed UV color is a function of an unkown
intrinsic UV color and a reasonably well known Galactic reddening.  The
relationship between the UV color excess $E(FUV-NUV)$ and the Galactic
extinction parameter $E(B-V)_{Gal}$ was modeled as a quadratic function
of the form.
\begin{equation}
E(FUV-NUV) = a\, E(B-V)_{Gal} + b\,E(B-V)_{Gal}^2
\end{equation}
where $a$ and $b$ are linear functions of the observed UV
color.  The same was done for the relationship between the FUV extinction
($A_{FUV}$) and $E(B-V)_{Gal}$. A range of solutions for the coefficients
$a$ and $b$ were determined by a least squares polynomial fit for each
curve of constant intrinsic reddening and burst age in the $E(FUV-NUV)$ -
$E(B-V)_{Gal}$ or $A_{FUV}$ - $E(B-V)_{Gal}$ plane. A linear fit was
then performed for each coeffecient as a function of observed UV color;
giving the final solutions of:
\begin{eqnarray*}
a_{E(FUV-NUV)}&=&0.78 + 0.155\,(FUV-NUV)\\
b_{E(FUV-NUV)}&=&0.559\, \\
a_{A_{FUV}}&=&8.749 - 0.077\,(FUV-NUV)\\
b_{A_{FUV}}&=&0.141 + 0.007\,(FUV-NUV),
\end{eqnarray*}
where $(FUV-NUV)$ is the UV color measured from the model spectra in AB 
magnitudes.
The technique correctly determines $A_{FUV}$ to within $\pm 0.03$
magnitudes and $E(FUV-NUV)$ to within $\pm 0.05$ magnitudes provided
that $E(B-V)_{Gal} \le 0.4$. The fit improves with decreasing
$E(B-V)_{Gal}$. A fit to $A_{NUV}$ was not as successful due to the
large width of the NUV bandpass. This method also fails for older
stellar populations, such as post-instantaneous burst with ages $>
300$\,Myr.  The net effect of this correction on our sample is
small. The average $A_{FUV}$ increases by 0.01 magnitudes and
$E(FUV-NUV)$ increases by 0.03 magnitudes.

\subsection{Calibrating STIS UV color and $\beta$}

We use the UV spectral slope $\beta$ defined by \citet{cal94} within
10 narrow windows between 1268 and 2580\,$\mbox{\AA}$. These windows avoid many
of the emision and absorption features in the UV.  
IUE spectral data for 33 starburst galaxies \citep{kin93} were
used in addition to 693 synthetic starburst spectra. The metallicity and
age parameters of the synthetic starburst spectra were described
previously; however, only models with continuous star formation rates
were used.

The IUE data were redshifted to $z=0.04$ with no further
processing. Intrinsic reddening was applied to the restframe starburst
spectra in accordance with the \citet{cal94} mean starburst
attenuation curve for $0 \le E(B-V)_{stars} \le 1$ and then redshifted
to $z=0.04$.

The broad band UV color $(FUV - NUV)$ was measured directly from these
spectra and fit to $\beta$ as defined above (and
measured in the rest-frame).  Comparing the UV color to $\beta$ reveals
the robust linear relationship: 
\begin{equation}
\beta = -2.20 + 1.88\,(FUV-NUV),
\end{equation}
determined from a minimized chi-square fit. Here, FUV and NUV are
measured in AB magnitudes. The fit is secure, with an rms scatter of
0.04 in $\beta$, over the range of STIS UV colors modeled
($-1.5<(FUV-NUV)<1.5$).

\clearpage

\figcaption{
We show the global
relationship between UV color and FIR/UV fluxes.  The line is a least-squares
fit to the IUE starburst sample of M99 (the dots), and is 
consistent with a shell
geometry for the dust distribution \citep{wit00}.
Included are the three galaxies from TKS; $\beta$ and IRX were
computed in a manner similar to our own galaxies, taking into account the
different wavelengths of their filters.  
Red leaks in the FOC near-UV
filter may be partially responsible for the great redness of VII Zw 031.
If that is true, then only the limit on the IRX (derived from the far-UV
data) would be reliable.
\label{global_betairx}}

\figcaption{The relationship between UV color and FIR/UV fluxes, in
2 kpc diameter apertures.  
The line is a simple foreground screen model fitted to the UV-selected
starbursts.  
The IR fluxes in
IRAS 08572+3915 and IRAS 19254--7245 have each been apportioned to 70\% for
the IR-bright nucleus and 30\% for the IR-faint nucleus, as discussed
earlier---note that these are conservative limits, and in both cases, 
the IR-bright nucleus may in fact
produce essentially all the far-IR luminosity.
\label{nuclear_betairx}}

\figcaption{Far-UV, near-UV, and near-IR views of 
VV 114, 10
kpc on a side.  The large tic marks are 1\arcsec\ apart.
\label{fig_vv114a}}

\figcaption{
The inner 3 kpc of VV 114 EAST in far-UV, near-UV, 
{\em F160W}, and 8.44 GHz.  The crosses mark the positions of the 
2.2 \micron\ peaks, sources A and B of \citet{doy95}.
The large tic marks are 1\arcsec\ apart.
\label{fig_vv114b}}

\figcaption{The inner 3 kpc of VV 114 WEST
at far-UV, near-UV, {\em F110W}, and {\em F160W}.
\label{fig_vv114c}}

\figcaption{Far-UV, near-UV, optical, and near-IR views of 
IRAS 08572+3915, 20
kpc on a side.  The large tic marks are 1\arcsec\ apart.
\label{fig_ir08572a}}

\figcaption{
The inner 3 kpc of IRAS 08572+3915 NW in far-UV, near-UV, {\em F814W}, and
{\em F160W}.  The cross mark the position of the peak at 2.2 \micron.
The large tic marks are 1\arcsec\ apart.
\label{fig_ir08572b}}

\figcaption{
The inner 3 kpc of IRAS 08572+3915 SE in far-UV, near-UV, {\em F814W}, and
{\em F160W}.  The cross mark the position of the peak at 2.2 \micron.
The large tic marks are 1\arcsec\ apart.
\label{fig_ir08572c}}

\figcaption{Far-UV, near-UV, and near-IR views of IC 883, 10
kpc on a side.  The large tic marks are 1\arcsec\ apart.
\label{fig_ic883a}}

\figcaption{
The inner 3 kpc of IC 883 in far-UV, near-UV, {\em F110W}, and 
in the radio at
8.44 GHz.  The large tic marks are 1\arcsec\ apart.  The crosses mark
the positions of the {\em F222M} peaks (see text).  Note that the astrometry
is not sufficiently accurate to align the radio image with the HST images,
so we have assumed that the brightest radio and 2.2 \micron\ peaks 
are coincident.
\label{fig_ic883b}}

\figcaption{a) Far-UV, near-UV, optical, and near-IR views of Mrk 273, 10
kpc on a side.  The large tic marks are 1\arcsec\ apart.  
\label{fig_mrk273a}}

\figcaption{
The inner 3 kpc of Mrk 273 in far-UV, near-UV, {\em F814W}, and
{\em F160W}.  The large tic marks are 1\arcsec\ apart.  The crosses mark
the positions of the radio/mid-IR peaks (see text).
\label{fig_mrk273b}}

\figcaption{ Far-UV, near-UV, optical, and near-IR views of 
IRAS 15250+3609, 10 kpc on a side.  The large tic marks are 1\arcsec\ apart.
\label{fig_ir15250a}}

\figcaption{
The inner 3 kpc of IRAS 15250+3609 in far-UV, near-UV, {\em F814W}, and
{\em F160W}.  The cross marks the position of the peak at 2.2 \micron.
The large tic marks are 1\arcsec\ apart.
\label{fig_ir15250b}}

\figcaption{a) Far-UV, near-UV, optical, and near-IR views of Arp 220, 10
kpc on a side.  The large tic marks are 1\arcsec\ apart.
\label{fig_arp220a}}

\figcaption{
The inner 3 kpc of Arp 220 in far-UV, near-UV, {\em F814W}, and
{\em F160W}.  The large tic marks are 1\arcsec\ apart.
\label{fig_arp220b}}

\figcaption{Far-UV, near-UV, and F814W views of IRAS 19254--7245, with a
10 kpc$\times$20 kpc field of view.  The large tic marks are 1\arcsec\ apart.
The crosses mark the positions of the brightest pixels in the {\em F814W} 
image.
\label{fig_ir19254a}}

\figcaption{
The inner 3 kpc of the northern and
southern nuclei of IRAS 19254--7245 in far-UV, near-UV and {\em F814W}.
The large tic marks are 1\arcsec\ apart.  The crosses mark the positions
of the nuclear peak pixels as measured in the {\em F814W} image.
\label{fig_ir19254b}}

\figcaption{We show the growth curves of flux from our galaxies.  The 
different point styles denote different nuclei.  The horizontal dashed lines
denote 100\%, 50\%, and 10\% of the flux contained within the 5 kpc 
radius aperture.  The curved dotted line shows the growth curve for a
source with constant surface brightness.  Arp 220 and IC 883, two galaxies
without any strong sources in the inner kpc, are closest to the dotted
line.  IRAS 08572+3915 SE and IRAS 15250+3609, two galaxies that do have
strong, centrally located sources, are farthest above the line, as would be
expected.  VV 114 E falls far below the line of constant surface
brightness, as it emits virtually no near-UV light of its own, and its
5 kpc radius flux is almost entirely due to spillover from VV 114 W, part of
which is included in the outer parts of the 5 kpc aperture centered on
VV 114 E.  The half-light radii in Table 4 were estimated using this
figure.
\label{halflight}}

\figcaption{We show the {\em F606W} surface brightnesses of our
galaxies, compared with the surface brightnesses of galaxies within
the WF chips in the HDF, if our galaxies were placed at (a) $z=1.54$; (b)
$z=3.12$.  The symbols are the same as in Fig. \ref{halflight}.
\label{sb_hdf}}

\figcaption{We show the {\em V-I} vs. {\em V} color-magnitude diagram
of our galaxies, if located at $z=3.12$, 
compared to the HDF {\em U}-band dropouts from M99.
The symbols are the same as in 
Fig. \ref{halflight}.
\label{udropouts_hdf}}
\clearpage

\begin{deluxetable}{lccccccc}
\tabletypesize{\scriptsize}
\tablecolumns{6}
\tablewidth{0pc}
\tablecaption{HST ULIG Sample and Physical Data\tablenotemark{a}}

\tablehead{
\colhead{Galaxy}&\colhead{redshift}&
\colhead{Linear scale\tablenotemark{b}}& 
\colhead{Luminosity dist.\tablenotemark{b}}&
\colhead{$A_{FUV}$}&\colhead{$E(FUV-NUV)$}&\colhead{log$(L_{IR}/L_{\odot})$}&
\colhead{FIR\tablenotemark{c}}\\
\colhead{}&\colhead{($z$)}&\colhead{(kpc/arcsec)}&
\colhead{(Mpc)}&
\colhead{(mag)}&\colhead{(mag)}&&\colhead{(W m$^{-2}$)}\\
\colhead{(1)}&\colhead{(2)}&
\colhead{(3)}&\colhead{(4)}&
\colhead{(5)}&\colhead{(6)}&
\colhead{(7)}&\colhead{(8)}
}
\startdata
VV 114&0.0203&0.41&88.7&0.146&0.015&11.72&$1.50\times10^{-12}$\\
IRAS 08572+3915&0.0583&1.13&260.4&0.245&0.023&12.17&$5.60\times10^{-13}$\\
IC 883&0.0234&0.47&102.0&0.114&0.012&11.61&$1.05\times10^{-12}$\\
Mrk 273&0.0382&0.76&168.2&0.076&0.008&12.17&$1.43\times10^{-12}$\\
IRAS 15250+3609&0.0557&1.08&248.3&0.174&0.018&12.05&$4.82\times10^{-13}$\\
Arp 220&0.0183&0.37&\ 79.4&0.463&0.051&12.18&$6.70\times10^{-12}$\\
IRAS 19254--7245&0.0611&1.18&273.4&0.785&0.085&12.10&$3.51\times10^{-13}$\\
\enddata
\tablenotetext{a}{Keys to columns follow.\\Col. 1---Sources, sorted in
order of increasing Right Ascension.\\Col. 2---redshift\\Col. 3---Linear
scale\\Col. 4---Luminosity distance\\Cols. 5-6---Milky Way far-UV extinction, 
and UV color excess, in magnitudes\\Col. 7---Infrared
luminosity\\Col. 8---Far-IR flux}
\tablenotetext{b}{For $H_0=70$ km/s/Mpc, $\Omega_{M}=0.3$, and 
$\Omega_{\Lambda}=0.7$}
\tablenotetext{c}{Sources for the FIR data are: Soifer et al. (1989), and 
Sanders et al. (1995) (for IRAS 19254--7245 only).}
\end{deluxetable}
\clearpage

\begin{deluxetable}{lccc}
\tablecolumns{4}
\tablewidth{0pc}
\tablecaption{Log of HST ULIG Observations}

\tablehead{
\colhead{Galaxy}&\colhead{Obs. date}&\colhead{FUV Exposure}&
\colhead{NUV Exposure}\\
\colhead{}&\colhead{(U.T.)}&\colhead{(sec)}&\colhead{(sec)}\\
\colhead{(1)}&\colhead{(2)}&
\colhead{(3)}&\colhead{(4)}
}

\startdata
VV 114&2000 Dec 25&3051&1500\\
IRAS 08572+3915&2000 Mar 1&3891&1500\\
IC 883&2000 Apr 5&3827&1500\\
Mrk 273&1999 May 13&3594&1500\\
IRAS 15250+3609&2000 May 27&3891&1500\\
Arp 220&2000 May 23&3749&1500\\
IRAS 19254--7245&1999 Aug 30&2829&1500\\
\enddata
\end{deluxetable}
\clearpage

\begin{deluxetable}{lrrrrrr}
\tabletypesize{\scriptsize}
\tablecolumns{7}
\tablewidth{0pc}
\tablecaption{HST ULIG Photometry\tablenotemark{a}}

\tablehead{
\colhead{}&\multicolumn{5}{c}{Aperture radius}&\colhead{Total\tablenotemark{b}}\\
\colhead{}&\colhead{0.5 kpc}&\colhead{1 kpc}&\colhead{2 kpc}&\colhead{3 kpc}&\colhead{5kpc}&\colhead{}\\
\colhead{(1)}&\colhead{(2)}&
\colhead{(3)}&\colhead{(4)}&
\colhead{(5)}&\colhead{(6)}&
\colhead{(7)}
}

\startdata
\cutinhead{VV 114E}
&1\farcs210&2\farcs420&4\farcs840&7\farcs260&12\farcs100&Total\tablenotemark{b}\\
\cline{2-7}\\
FUV&22.70&21.08&19.21&17.70&15.60&14.73\\
unc.&0.23&0.21&0.15&0.08&0.03&0.02\\
NUV&22.10&20.39&18.38&17.00&15.08&14.26\\
unc.&0.09&0.07&0.04&0.03&0.01&0.01\\
$\beta$&$-1.12$&$-0.93$&$-0.67$&$-0.92$&$-1.25$&$-1.35$\\
unc.&0.46&0.41&0.29&0.16&0.07&0.04\\
log(IRX)&4.10&3.45&2.71&2.11&1.27&1.13\\
\\
\cutinhead{VV 114W}
&1\farcs210&2\farcs420&4\farcs840&7\farcs260&12\farcs100&Total\tablenotemark{b}\\
\cline{2-7}\\
FUV&17.70&16.58&15.71&14.98&14.78&\nodata\\
unc.&0.01&0.01&0.01&0.01&0.02&\nodata\\
NUV&17.14&16.02&15.20&14.53&14.29&\nodata\\
unc.&0.01&0.01&0.01&0.01&0.01&\nodata\\
$\beta$&$-1.18$&$-1.18$&$-1.28$&$-1.39$&$-1.33$&\nodata\\
unc.&0.01&0.01&0.01&0.01&0.03&\nodata\\
log(IRX)&1.92&1.48&1.13&0.84&0.76&\nodata\\
\\
\cutinhead{IRAS 08572+3915NW}
&0\farcs444&0\farcs887&1\farcs774&2\farcs661&4\farcs435&Total\tablenotemark{b}\\
\cline{2-7}\\
FUV&25.90&23.60&22.02&21.21&20.39&19.24\\
unc.&0.33&0.16&0.14&0.15&0.20&0.27\\
NUV&24.88&22.52&21.08&20.41&19.82&18.99\\
unc.&0.21&0.10&0.10&0.12&0.20&0.18\\
$\beta$&$-0.31$&$-0.21$&$-0.47$&$-0.72$&$-1.16$&$-0.66$\\
unc.&0.74&0.34&0.33&0.37&0.53&0.61\\
log(IRX)&4.98&4.06&3.43&3.11&2.78&2.47\\
\\
\cutinhead{IRAS 08572+3915SE}
&0\farcs444&0\farcs887&1\farcs774&2\farcs661&4\farcs435&Total\tablenotemark{b}\\
\cline{2-7}\\
FUV&22.15&21.67&21.14&20.74&19.91&\nodata\\
unc.&0.02&0.03&0.06&0.10&0.13&\nodata\\
NUV&21.41&21.01&20.64&20.44&19.82&\nodata\\
unc.&0.01&0.03&0.07&0.13&0.20&\nodata\\
$\beta$&$-0.85$&$-1.01$&$-1.31$&$-1.66$&$-2.07$&\nodata\\
unc.&0.04&0.07&0.18&0.30&0.45&\nodata\\
log(IRX)&3.12&2.92&2.71&2.55&2.22\\
\\
\cutinhead{IC 883}
&1\farcs059&2\farcs117&4\farcs235&6\farcs352&10\farcs587&Total\tablenotemark{b}\\
\cline{2-7}\\
FUV&22.82&21.26&19.80&19.12&18.58&18.45\\
unc.&0.09&0.09&0.09&0.11&0.18&0.16\\
NUV&21.72&20.15&18.75&18.14&17.71&17.61\\
unc.&0.04&0.04&0.04&0.05&0.10&0.15\\
$\beta$&$-0.17$&$-0.14$&$-0.25$&$-0.39$&$-0.58$&$-0.64$\\
unc.&0.19&0.18&0.18&0.22&0.39&0.41\\
log(IRX)&4.23&3.61&3.02&2.75&2.54&2.48\\
\\
\cutinhead{Mrk 273}
&0\farcs661&1\farcs322&2\farcs644&3\farcs966&6\farcs610&Total\tablenotemark{b}\\
\cline{2-7}\\
FUV&22.37&21.39&20.48&19.79&19.00&18.66\\
unc.&0.02&0.03&0.05&0.07&0.09&0.12\\
NUV&20.84&19.81&19.07&18.56&17.96&17.78\\
unc.&0.01&0.02&0.03&0.05&0.08&0.13\\
$\beta$&0.65&0.76&0.43&0.10&$-0.26$&$-0.56$\\
unc.&0.05&0.07&0.12&0.15&0.22&0.33\\
log(IRX)&4.20&3.81&3.44&3.17&2.85&2.71\\
\\
\cutinhead{IRAS 15250+3609}
&0\farcs463&0\farcs926&1\farcs852&2\farcs778&4\farcs630&Total\tablenotemark{b}\\
\cline{2-7}\\
FUV&21.81&19.88&19.30&19.07&18.93&18.73\\
unc.&0.01&0.01&0.01&0.02&0.05&0.10\\
NUV&20.85&19.03&18.47&18.23&18.11&18.01\\
unc.&0.01&0.01&0.01&0.01&0.03&0.06\\
$\beta$&$-0.43$&$-0.63$&$-0.68$&$-0.67$&$-0.71$&$-0.87$\\
unc.&0.03&0.01&0.03&0.05&0.11&0.22\\
log(IRX)&3.47&2.69&2.46&2.37&2.31&2.23\\
\\
\cutinhead{Arp 220}
&1\farcs346&2\farcs692&5\farcs384&8\farcs076&13\farcs460&Total\tablenotemark{b}\\*
\cline{2-7}\\
FUV&24.30&22.83&21.03&20.08&19.73&19.68\\
unc.&0.47&0.49&0.37&0.35&0.69&0.31\\
NUV&23.84&22.41&20.30&19.22&18.52&18.51\\
unc.&0.46&0.50&0.28&0.24&0.35&0.14\\
$\beta$&$-1.44$&$-1.51$&$-0.94$&$-0.67$&$-0.04$&$-0.09$\\
unc.&1.24&1.30&0.87&0.79&1.45&0.64\\
log(IRX)&5.49&4.9&4.18&3.80&3.66&3.64\\
\cutinhead{IRAS 19254--7245N}
&0\farcs425&0\farcs849&1\farcs698&2\farcs547&4\farcs245&Total\tablenotemark{b}\\
\cline{2-7}\\
FUV&25.43&24.08&23.14&22.31&21.44&\nodata\\
unc.&0.12&0.12&0.20&0.21&0.26&\nodata\\
NUV&23.70&22.72&22.10&21.56&20.85&\nodata\\
unc.&0.07&0.11&0.23&0.32&0.46&\nodata\\
$\beta$&0.89&0.19&$-0.40$&$-0.95$&$-1.25$&\nodata\\
unc.&0.27&0.31&0.58&0.72&1.00&\nodata\\
log(IRX)&4.01&3.47&3.09&2.76&2.41\\
\cutinhead{IRAS 19254--7245S}
&0\farcs425&0\farcs849&1\farcs698&2\farcs547&4\farcs245&Total\tablenotemark{b}\\
\cline{2-7}\\
FUV&$>$27.24&$>$25.73&23.87&22.60&21.74&20.31\\
unc.&\nodata&\nodata&0.39&0.27&0.34&0.25\\
NUV&23.64&23.17&22.20&21.34&20.59&19.58\\
unc.&0.07&0.16&0.26&0.26&0.36&0.44\\
$\beta$&$>4.41$&$>3.38$&0.77&0.00&$-0.20$&$-0.98$\\
unc.&\nodata&\nodata&0.88&0.71&0.94&0.95\\
log(IRX)&$>5.10$&$>4.70$&3.75&3.25&2.90&2.48\\
\enddata
\tablenotetext{a}{Photometry is given in the AB magnitude system.  Upper
limits are $2\sigma$.}
\tablenotetext{b}{Quantities include the contributions from 
both components of the system.}
\end{deluxetable}
\clearpage

\begin{deluxetable}{lrcc}
\tablecolumns{9}
\tablewidth{0pc}
\tablecaption{Predictions for detecting ULIGs in the HDF}
\tablehead{
\colhead{Galaxy}&\colhead{$R_{0.5}$}&
\multicolumn{2}{c}{Surface brightness ($F606W_{AB}$})\\
\cline{3-4}\\
\colhead{}&\colhead{}&\colhead{$z=1.54$}&\colhead{$z=3.12$}\\
\colhead{}&\colhead{(kpc)}&\colhead{(Mag/arcsec$^2$)}&
\colhead{(Mag/arcsec$^2$)}\\
\colhead{(1)}&\colhead{(2)}&
\colhead{(3)}&\colhead{(4)}}

\startdata
VV 114 E&4&24.84&26.93\\
VV 114 W&2.25&22.81&24.85\\
IRAS 08572+3915NW&2.75&26.37&28.50\\
IRAS 08572+3915SE&2.25&25.94&27.59\\
IC 883&2.75&26.39&28.82\\
Mrk 273&2.75&25.60&28.21\\
IR 15250+3609&1.25&23.13&25.49\\
Arp 220&3&27.62&30.35\\
IR 19254--7245N&3&27.02&29.10\\
IR 19254--7245S&3&26.76&29.39\\
\enddata
\end{deluxetable}
\clearpage

\begin{deluxetable}{lcc}
\tablecolumns{3}
\tablewidth{0pc}
\tablecaption{Absolute Far-UV magnitudes for ULIGs\tablenotemark{a}}
\tablehead{
\colhead{Galaxy}&\colhead{$M_{FUV}$ (MW ext. only)\tablenotemark{b}}&
\colhead{$M_{FUV}$ (MW+Int. ext.)\tablenotemark{b}}\\
\colhead{}&\colhead{(Mag.)}&\colhead{(Mag)}\\
\colhead{(1)}&\colhead{(2)}&\colhead{(3)}}

\startdata
VV 114 E&$-19.27$&$-21.2$\\
VV 114 W&$-20.09$&$-21.8$\\
IRAS 08572+3915NW&$-16.88$&$-19.0$\\
IRAS 08572+3915SE&$-17.36$&$-17.7$\\
IC 883&$-16.54$&$-19.7$\\
Mrk 273&$-17.17$&$-21.1$\\
IR 15250+3609&$-18.15$&$-21.2$\\
Arp 220&$-15.21$&$-19.5$\\
IR 19254--7245N&$-16.47$&$-18.4$\\
IR 19254--7245S&$-16.17$&$-20.2$\\
\enddata
\tablenotetext{a}{UV photometry is in 5 kpc physical apertures. 
Photometry is given in the AB system.}
\tablenotetext{b}{Column (2) gives absolute magnitudes corrected
for foreground (Milky Way) extinction only.  Column (3) also includes a
correction for internal extinction, $A_{FUV}\approx 4.4+2 \beta$,
taken from M99.} 
\end{deluxetable}
\clearpage

\begin{deluxetable}{lcccccc}
\tablecolumns{7}
\tablewidth{0pc}
\tablecaption{Predicted colors for ULIGs at high redshifts\tablenotemark{a}}
\tablehead{
\colhead{Galaxy}&
\multicolumn{3}{c}{At $z=1.96$}&
\multicolumn{3}{c}{At $z=3.80$}\\
\cline{2-4} \cline{5-7}\\
\colhead{}&\colhead{$R$}&\colhead{$K$}&\colhead{$R-K$}&
\colhead{$R$}&\colhead{$K$}&\colhead{$R-K$}\\
\colhead{}&\colhead{(Mag)}&\colhead{(Mag)}&\colhead{(Mag)}&\colhead{(Mag)}&
\colhead{(Mag)}&\colhead{(Mag)}\\
\colhead{(1)}&\colhead{(2)}&
\colhead{(3)}&\colhead{(4)}&
\colhead{(5)}&\colhead{(6)}&
\colhead{(7)}
}

\startdata
IRAS 08572+3915NW&27.01&20.82&6.19&28.78&23.11&5.67\\
Mrk 273&26.24&21.59&4.65&28.49&23.61&4.88\\
IRAS 15250+3609&25.48&22.33&3.15&27.48&24.12&2.66\\
Arp 220&28.07&22.52&5.55&30.44&24.49&5.95\\
IRAS 19254--7245S&27.21&21.69&5.52&29.48&24.10&5.38\\
\enddata
\tablenotetext{a}{Predicted using UV photometry is in 5 kpc physical apertures, 
and optical photometry through 10\arcsec\ diameter apertures, except for
IRAS 19254--7245 (8\arcsec\ optical).  Photometry is given in the Johnson
system in this table only.}
\end{deluxetable}

\begin{thebibliography}{}
\bibitem[Adelberger \& Steidel (2000)]{ade00} Adelberger, K. L., \& 
Steidel, C. C. 2000, \apj, 544, 218

\bibitem[Aussel et al. (1999)]{aus99} Aussel, H., Cesarsky, C. J.,
Elbaz, D. \& Starck, J. L. 1999, \aap, 342, 313

\bibitem[Barger et al. (1999a)]{bar99a}Barger, A. J., Cowie, L. L., Smail, I., 
Ivison, R. J., Blain, A. W., \& Kneib, J.-P. 1999a, \apj, 117, 2656

\bibitem[Barger, Cowie, \& Sanders (1999)]{bar99b} Barger, A. J., Cowie,
L. L., \& Sanders, D. B. 1999, \apjl, 518, L5

\bibitem[Barth \& Shields (2000)]{bar00} Barth, A. J., \& Shields, J. C.
2000, \pasp, 112, 753

\bibitem[Bryant \& Scoville (1999)]{bry99} Bryant, P. M., \& Scoville, N. Z.
1999, \aj, 117, 2632

\bibitem[Calzetti, Kinney \& Storchi-Bergmann (1994)]{cal94} Calzetti, D.,
Kinney, A. L., \& Storchi-Bergmann, T. 1994, \apj, 429, 582

\bibitem[Cardelli, Clayton \& Mathis (1989)]{car89} Cardelli, J. A.,
Clayton, G. C. \& Mathis, J. S. 1989, \apj, 345, 245

\bibitem[Carilli \& Taylor (2000)]{car00} Carilli, C. L., \& Taylor, G. B.
2000, \apjl, 532, L95

\bibitem[Carroll, Press \& Turner (1992)]{car92} Carroll, S. M., \& Press, W. H.,
\& Turner, E. L. 1992, \araa, 30, 499

\bibitem[Chapman et al. (2000)]{cha00} Chapman, S. C. et al. 2000,
\mnras, 319, 318

\bibitem[Condon et al. (1990)]{con90} Condon, J. J., Helou, G., Sanders,
D. B., \& Soifer, B. T. 1990, \apjs, 73, 359

\bibitem[Condon et al. (1991)]{con91} Condon, J. J., Huang, Z.-P.,
Yin, Q. F., \& Thuan, T. X. 1991, \apj, 378, 65

\bibitem[de Grijp et al. (1985)]{deg85} de Grijp, M. H. K., Lub, J., \&
de Jong, T. 1985, \nat, 314, 240

\bibitem[Dey et al. (1999)]{dey99} Dey, A. et al. (1999), \apj, 519, 610

\bibitem[Downes \& Solomon (1998)]{dow98} Downes, D., \& Solomon, P. M.
1998, \apj, 507, 615

\bibitem[Doyon et al. (1995)]{doy95} Doyon, R.,  Nadeau, D.,
Joseph, R. D., Goldader, J. D., Sanders, D. B., \& Rowlands, N.
1995, \apj, 450, 111

\bibitem[Dudley (1999)]{dud99} Dudley, C. C. 1999, \mnras, 307, 553

\bibitem[Dudley \& Wynn-Williams (1997)]{dud97} Dudley, C. C. \&
Wynn-Williams, C. G. 1997, \apj, 488, 720

\bibitem[Frayer et al. (1999)]{fra99} Frayer, D. T., Ivison, R. J.,
Smail, I., Yun, M. S., \& Armus, L. 1999, \apj, 118, 139

\bibitem[Frayer et al. (2000)]{fra00} Frayer, D. T., Smail, I., Ivison,
R. J., \& Scoville, N. Z. 2000, \aj, 120, 1668

\bibitem[Genzel et al. (1998)]{gen98} Genzel, R. et al. 1998,
\apj, 498, 579

\bibitem[Goldader et al. (1995)]{gol95} Goldader, J. D., Joseph,
R. D., Doyon, R., \& Sanders, D. B. 1995, \apj, 444, 97

\bibitem[Goldader et al. (1997a)]{gol97a} Goldader, J. D., Joseph,
R. D., Doyon, R., \& Sanders, D. B. 1997a, \apjs, 108, 449

\bibitem[Goldader et al. (1997b)]{gol97b} Goldader, J. D., 
Goldader, D. L., Joseph, R. D., Doyon, R., \& Sanders, D. B. 1997b,
\aj, 113, 1569

\bibitem[Gonz\'{a}lez Delgado, Heckman, \& Leitherer (2001)]{gon01}
Gonz\'{a}lez Delgado, R. M., Heckman, T. M., \& Leitherer, C. 2001,
\apj, 546, 845

\bibitem[Graham et al. (1990)]{gra90} Graham, J. R., Carico, D. P.,
Matthews, K., Neugebauer, G., Soifer, B. T., \& Wilson, T. D. 1990,
\apjl, 354, L5

\bibitem[Hauser et al. (1998)]{hau98} Hauser, M. et al. 1998, \apj, 508, 25

\bibitem[Heckman, Armus \& Miley (1987)]{hec87} Heckman, T. M.,
Armus, L., \& Miley, G. K. 1987, \aj, 92, 276

\bibitem[Heckman et al. (1998)]{hec98} Heckman, T. M., Robert, C., Leitherer,
C., Garnett, D. R \& vsd der Rydt,, F. 1998, \apj, 503, 646

\bibitem[Helou et al. (1988)]{hel88} Helou, G., Khan, I. R., Malek, L.,
\& Boehmer, L. 1988, \apjs, 68, 151

\bibitem[Ho (1999)]{ho99}Ho, L. C. 1999, Advances in Space Research, 23, 813

\bibitem[Holland et al. (1999)]{hol99} Holland, W. S., et al. 1999, \mnras
303, 659

\bibitem[Ivison et al. (1998)]{ivi98} Ivison, R. J., et al. 1998, \mnras, 
298, 583

\bibitem[Ivison et al. (2000)]{ivi00a} Ivison, R. J., et al. 2000,
\apj, 542, 27 (2000a)

\bibitem[Ivison et al. (2000)]{ivi00b} Ivison, R. J., et al. 2000,
\mnras, 315, 209 (2000b)

\bibitem[Ivison et al. (2001)]{ivi01} Ivison, R., Smail, I., Frayer, D.,
Kneib, J.-P. \& Blain, A. 2001, \apjl, in press (astro-ph/0110085)

\bibitem[Iwasawa (1999)]{iwa99} Iwasawa, K. 1999, \mnras, 302, 96

\bibitem[Kim, Veilleux \& Sanders (1998)]{kim98} Kim, D.-C., Veilleux, S.
\& Sanders, D. B. 1998, \apj, 508, 627

\bibitem[Kinney et al. (1993)]{kin93} Kinney, A. L., Bohlin, R. C.,
Calzetti, D., Panagia, N. \& Wyse, R. F. G. 1993, \apjs, 86, 5

\bibitem[Knapen et al. (1997)]{kna97} Knapen, J. H. et al. 1997,
\apjl, 490, L29 

\bibitem[Knop et al. (1994)]{kno94} Knop, R. A., Soifer, B. T., Graham, J. R.,
Matthews, K., Sanders, D. B., \& Scoville, N. Z. 1994, \aj, 107, 920

\bibitem[Laurent et al. (2000a)]{lau00a} Laurent, O., et al.
2000a, \aap, 359, 887

\bibitem[Laurent et al. (2000b)]{lau00b} Laurent, O., et al.
2000b, \aap, in preparation

\bibitem[Leitherer et al. (1999)]{lei99} Leitherer, C., et al. 1999,
\apjs, 123, 3

\bibitem[Lutz, Veilleux \& Genzel (1999)]{lut99} Lutz, D., Veilleux, S., \& 
Genzel, R. 1999, \apj, 517, L13

\bibitem[Madau et al. (1996)]{mad96} Madau, P., Ferguson, H. C.,
Dickinson, M. E., Giavalisco, M., Steidel, C. C. \& Fruchter, A. 1996,
\mnras, 283, 1388

\bibitem[Meurer et al. (1995)]{meu95} Meurer, G. R., Heckman, T. M., 
Leitherer, C., Kinney, A., Robert, C., \& Garnett, D. R. 1995,
\aj, 110, 2665

\bibitem[Meurer et al. (1997)]{meu97} Meurer, G. R., Heckman, T. M., 
Lehnert, M. D., Leitherer, C., \& Lowenthal, J. 1997, \aj, 114, 54

\bibitem[Meurer et al. (1999)]{meu99} Meurer, G. R., Heckman, T. M., 
\& Calzetti, D. 1999, \apj, 521, 64 (M99)

\bibitem[Mirabel, Lutz \& Maza (1991)]{mir91} Mirabel, I. F.,
Lutz, D., \& Maza, J. 1991, \aap, 243, 367

\bibitem[Risaliti et al. (2000)]{ris00} Risaliti, G., Gilli, R.,
Maiolino, R., \& Salvati, M. 2000, \aap, 357, 13

\bibitem[Sanders et al. (1988)]{san88} Sanders, D. B. et al.
1988, \apj, 325, 74

\bibitem[Sanders et al. (1991)]{san91} Sanders, D. B., Scoville,
N. Z., \& Soifer, B. T. 1991, \apj, 370, 158

\bibitem[Sanders et al. (1995)]{san95} Sanders, D. B., Egami, E.,
Lipari, S., Mirabel, I. F., \& Soifer, B. T. 1995, \aj, 110, 1993

\bibitem[Sanders \& Mirabel (1996)]{san96} Sanders, D. B., \& Mirabel,
I. F. 1996, \araa, 34, 749

\bibitem[Scoville et al. (1991)]{sco91} Scoville, N. Z., Sargent, A. I.,
Sanders, D. B., \& Soifer, B. T. 1991, \apjl, 366, 5

\bibitem[Scoville et al. (2000)]{sco00} Scoville, N. Z., et al.
2000, \aj, 119, 991

\bibitem[Smail et al. (1999)]{sma99} Smail, I., et al. 1999, \mnras, 308, 
1061

\bibitem[Smail et al. (2000)]{sma00} Smail, I., Ivison, R., Blain, A.,
\& Kneib, J.-P. 2000, to appear in proceedings of U. Massachusetts/INAOE
Conference ``Deep Millimeter Surveys"

\bibitem[Smith et al. (1998)]{smi98} Smith, H. E., Lonsdale, C. J.,
\& Lonsdale, C. J. 1998, \apj, 492, 137

\bibitem[Soifer et al. (1987)]{soi87} Soifer, B. T., et al. 1987,
\apj, 320, 238

\bibitem[Soifer et al. (1989)]{soi89} Soifer, B. T., Boehmer, L.,
Neugebauer, G., \& Sanders, D. B. 1989, \aj, 98, 766

\bibitem[Soifer et al. (2000)]{soi00} Soifer, B. T., et al. 2000,
\aj, 119, 509

\bibitem[Soifer et al. (2001)]{soi01} Soifer, B. T., et al. 2001,
\aj, in press (astro-ph/0106172)

\bibitem[Surace et al. (1998)]{sur98} Surace, J. A., Sanders,
D. B., Vacca, W. D., Veilleux, S., \& Mazzarella, J. M. 1998, \apj, 492, 116

\bibitem[Surace \& Sanders (2000)]{sur00a} Surace, J. A., \& Sanders,
D. B. 2000, \aj, 120, 604

\bibitem[Thuan \& Gunn (1976)]{thu76} Thuan, T. X., \& Gunn, J. E. 1976, 
\pasp, 88, 543

\bibitem[Totani \& Yoshi (2000)]{tot00} Totani, T., \& Yoshi, Y. 2000, 
\apj, 540, 81

\bibitem[Trentham, Kormendy, \& Sanders (1999)]{tre99} Trentham, N.,
Kormendy, J. K., \& Sanders, D. B. 1999, \aj, 117, 2152 (TKS)

\bibitem[Veilleux, Kim \& Sanders (1995)]{vei95} Veilleux, S., Kim, D.-C.,
Sanders, D. B., Mazzarella, J. M., \& Soifer, B. T. 1995, \apjs, 98, 171

\bibitem[Veilleux, Kim \& Sanders (1999)]{vei99} Veilleux, S., Kim, D.-C.,
and Sanders, D. B. 1999, \apj, 522, 113

\bibitem[Wade et al. (1979)]{wad79} Wade, R. A., Hoessel, J. G., Elias,
J. H., \& Huchra, J. P. 1979, \pasp, 91, 35

\bibitem[Williams et al. (1996)]{wil96} Williams, R., et al. 1996,
\aj, 112, 1335

\bibitem[Witt \& Gordon (2000)]{wit00} Witt, A. N. \& Gordon, K. D. 2000,
\apj, 528, 799

\bibitem[Xia et al. (2001)]{xia01} Xia, X.-Y., Xue, S. J., Mao, S., Boller, 
Th., Deng, Z. G. \& Wu, H. 2001, \apj, in press (astro-ph/0107559)

\bibitem[Yun, Scoville \& Knop (1994)]{yun94} Yun, M. S., Scoville, N. Z.,
\& Knop, R. A. 1994, \apjl, 430, L109

\end{thebibliography}
\end{document}